\documentclass[aps,preprint,reprint,nofootinbib,superscriptaddress]{revtex4}

\usepackage{graphicx}
\usepackage{xcolor}
\usepackage{amssymb}
\usepackage{amsmath}
\usepackage{bm}
\usepackage{hyperref}

\usepackage{datetime}
\usepackage[T1]{fontenc}
\usepackage{ulem}

\makeatletter
\renewcommand\tableofcontents{%
    \begingroup
    {\Large \bfseries Contents\par}
    \@starttoc{toc}
    \endgroup
}
\makeatother

\newcommand{\orcid}[1]{\,\href{https://orcid.org/#1}{\includegraphics[width=9pt]{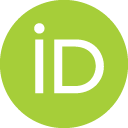}}\,}

\newcommand{\nslash}{n\kern -0.50em /}
\newcommand{\Sslash}{\kern 0.2 em S\kern -0.50em /}

\begin{document}

\title{Systematic uncertainties from higher-twist corrections \\ in DIS at large $x$}

\author{Matteo Cerutti\orcid{\orcidMC}}
\email{mcerutti@jlab.org}
\affiliation{Christopher Newport University, Newport News, Virginia 23606, USA}
\affiliation{Jefferson Lab, Newport News, Virginia 23606, USA}
\newcommand{\orcidMC}{0000-0001-7238-5657}

\author{Alberto Accardi\orcid{\orcidAA}}
\email{accardi@jlab.org}
\affiliation{Christopher Newport University, Newport News, Virginia 23606, USA}
\affiliation{Jefferson Lab, Newport News, Virginia 23606, USA}
\newcommand{\orcidAA}{0000-0002-2077-6557}

\author{I. P. Fernando\orcid{\orcidIPF}}
\email{ishara@virginia.edu}
\affiliation{University of Virginia, Charlottesville, Virginia 22904, USA}
\newcommand{\orcidIPF}{0000-0002-7570-7778}

\author{Shujie Li\orcid{\orcidSL}}
\email{shujieli@lbl.gov}
\affiliation{University of New Hampshire, Durham, New Hampshire 03824, USA}
\affiliation{Lawrence Berkeley National Laboratory, Berkeley, California 94720, USA}
\newcommand{\orcidSL}{0000-0003-1252-5392}

\author{J.F. Owens\orcid{\orcidJO}}
\email{owens@hep.fsu.edu}
\affiliation{Florida State University, Tallahassee, Florida 32306, USA \\ \ }
\newcommand{\orcidJO}{0000-0002-7351-0218}

\author{Sanghwa Park\orcid{\orcidSP}}
\email{sanghwa@jlab.org}
\affiliation{Jefferson Lab, Newport News, Virginia 23606, USA}
\newcommand{\orcidSP}{0000-0002-8898-1231}

\begin{abstract}
We investigate the systematic uncertainties and potential biases arising from the inclusion of large-$x$ corrections to proton and deuteron deep inelastic scattering (DIS) data in global quantum chromodynamics (QCD) analyses. Using the CTEQ-JLab framework, we examine various approaches to implementing higher-twist corrections in nucleon structure functions and off-shell PDF modifications in deuteron targets. We analyze how these components interact and influence the determination of the $d$-quark PDF and the neutron structure function at large $x$. We find that it is very important to consider isospin-dependent higher-twist corrections in order to minimize implementation biases in the extracted quantities.
\end{abstract}

\preprint{JLAB-THY-25-7}

\maketitle
\tableofcontents
\newpage

\section{Introduction}
Knowledge of parton distribution functions (PDFs) at large longitudinal momentum fraction $x$ is crucial to characterize the effects of color confinement and nonperturbative interactions on the proton's structure \cite{Brodsky:1994kg,Melnitchouk:1995fc,Isgur:1998yb,Holt:2010vj,Roberts:2013mja} and to search for beyond the Standard Model effects, \textit{e.g.}, in large-mass and forward-particle production at the Large Hadron Collider \cite{Accardi:2015lcq,Accardi:2016ndt,Accardi:2021ysh,Ball:2022qtp,Hammou:2023heg,Hammou:2024xuj,Ubiali:2024pyg} or in very precise measurements of parity-violating processes at Jefferson Lab~\cite{PVDIS:2014cmd,JeffersonLabSoLID:2022iod,Boughezal:2021kla,Bacchetta:2023hlw}.

In a global QCD analysis, such as from the CTEQ-JLab (CJ) collaboration \cite{Accardi:2009br,Owens:2012bv,Accardi:2016qay,Accardi:2023gyr}, measurements from different experiments can be included to constrain the large-$x$ behavior of the PDFs flavor by flavor \cite{Jimenez-Delgado:2013sma,Kovarik:2019xvh}. For the $u$-quark, information can be gathered from the large amount of Deep-Inelastic Scattering (DIS) data on proton targets, from fixed-target Drell-Yan data, and from jet production in hadron-hadron collisions.
For the $d$-quark, strong constraints come from precision data on large rapidity $W$-boson asymmetries in proton-antiproton collisions, and from abundant inclusive DIS data on deuteron targets, as well as from pioneering proton-tagged deuteron DIS at JLab.  
An accurate description of the deuteron DIS data, which are predominantly taken at small energy scale $Q^2$, requires correcting the theoretical calculations to account for the nuclear dynamics of the target nucleons (including nuclear binding, Fermi motion, and nucleon off-shell deformations) and for $1/Q^2$ power-suppressed effects. The latter include $O(M^2/Q^2)$ kinematic target mass corrections (TMCs), $O(\Lambda_\text{QCD}^2/Q^2)$ dynamical effects such as multiparton correlations, and any other ``residual'' $1/Q^2$ corrections, from, \textit{e.g.} higher-order QCD diagrams \cite{Blumlein:2006be,Blumlein:2008kz,Blumlein:2012se} or implementation choices of target mass corrections \cite{Accardi:2009br}. Although TMCs can be calculated \cite{Georgi:1976ve,DeRujula:1976baf,Aivazis:1993kh,Kretzer:2002fr,Accardi:2008ne,Steffens:2012jx,Schienbein:2007gr,Brady:2011uy,Ruiz:2023ozv}, the remaining terms are generally fitted to the data and collectively named ``higher-twist (HT) corrections''. 
\cite{Virchaux:1991jc,Alekhin:2003qq,Blumlein:2006be,Alekhin:2008ua,Blumlein:2008kz,Accardi:2009br,Owens:2012bv,Blumlein:2012se,Accardi:2016qay,Accardi:2023gyr,Alekhin:2012ig,Cocuzza:2021rfn,Alekhin:2017kpj,Alekhin:2017fpf,Alekhin:2022tip,Alekhin:2022uwc}.
Likewise, regarding nuclear dynamics, Fermi motion and binding effects can be calculated by means of non-relativistic nuclear wave functions \cite{Kulagin:1989mu,Kulagin:1994fz,Kulagin:2004ie,Kulagin:2007ph,Kahn:2008nq,DelDotto:2016vkh,Fornetti:2023gvf}. In a global QCD fit one can then leverage free-proton target data such as the above-mentioned $W$-boson rapidity asymmetry to fit the free-nucleon $d/u$ ratio, and utilize deuteron DIS measurements to constrain the off-shell deformations in nuclear targets.

In this paper, we focus on the interplay of the fitted off-shell and higher-twist corrections. We show that the implementation choices for HT corrections may introduce a bias in the calculation of the large-$x$ behavior of the DIS structure functions, and argue that this bias can be, in turn, partially but artificially compensated for by the fitted off-shell effects. Working consistently in the CJ22 framework \cite{Accardi:2023gyr} we demonstrate the presence of such bias in an actual PDF fit, identify safe implementations of the HT corrections, and discuss the remaining systematic uncertainty. We also briefly comment on similarities with studies of off-shell nucleon deformations performed by Alekhin, Kulagin and Petti (AKP) \cite{Alekhin:2022tip,Alekhin:2022uwc} and by the JAM collaboration \cite{Cocuzza:2021rfn}.

An interesting aspect of the interplay between HT corrections and off-shell effects is their different dependence on $Q^2$. As $Q^2$ increases, the HT corrections decrease, while the off-shell effects 
remain largely $Q^2$ independent. This suggests that the interconnection discussed here may differ for higher-$Q^2$ 
data, offering a cleaner opportunity to constrain off-shell effects when new data on proton and deuteron target DIS become available, \textit{e.g.}, from the future JLab 22 GeV upgrade~\cite{Accardi:2023chb}.

The plan of the paper is as follows. The QCD framework used in this analysis is described in Sec.~\ref{s:Formalism}. The comparison of fits obtained with various parametrizations is discussed in Sec.~\ref{s:results}. In Sec.~\ref{s:noW}, we discuss the impact of the $W$-boson asymmetry data on our results. Our conclusions are presented in Sec.~\ref{s:conclusion} and various technical details are reported in a series of appendices.

\section{Global QCD analysis framework}
\label{s:Formalism}

In this section, we outline the key theoretical ingredients needed to calculate the experimental observables considered in this analysis, including: the PDF parametrizations and QCD setup, the treatment of the bound nucleon structure function in the deuteron, and the functional form for the $1/Q^2$ corrections. We then examine the interplay of theoretical assumptions in the global fit, identify biases in the implementation of HT corrections, which are particularly significant at large $x$, and present our approach to mitigate these.

\subsection{PDF parametrizations and QCD setup}

The fits reported in this work rely on the  PDF parametrizations adopted in the latest CTEQ-JLab global fit (CJ22)~\cite{Accardi:2023gyr}. In particular, we use a standard five-parameter functional form on the initial scale $Q_0^2 = 1.69$ GeV$^2$ for most of the parton species, $\phi(x) = a_0 x^{a_1} (1-x)^{a_2} (1 + a_3\sqrt x + a_4 x)$ with $\phi=u_v, \, d_v, \, \bar d + \bar u, \, \bar d - \bar u, \, g$. Since the global dataset included in our analyses imposes little constraints on the strange and heavier quarks, we set $s=\bar s=0.4(\bar d + \bar u)$, and we consider the charm and bottom quarks as generated perturbatively. In order to allow the $d/u$ ratio to have a finite limit as $x$ approaches 1, we mix it with the $u_v$ quark at the initial scale:
\begin{equation}
    \label{e:PDF_du}
d_v^{\,\text{CJ}} (x, Q_0^2) = a_0^{d_v} \bigg ( \frac{d_v (x, Q_0^2)}{a_0^{d_v}} + b x^c u_v (x, Q_0^2) \bigg ).
\end{equation}
The normalization $a_0^{d_v}$ is determined by the valence sum rule, $b$ is a free parameter, and $c = 2$ is fixed due to the lack of strong constraints from the currently available experimental data.

The theoretical setup is also the same as in the CJ22 fit. More specifically, we calculate observables at next-to-leading order (NLO) perturbative accuracy, and for DIS observables we apply the 
ACOT-$\chi$ heavy-quark scheme~\cite{Kramer:2000hn}. Target-mass corrections, nuclear corrections, and other power corrections will be discussed in detail in the following subsections.
For computational speed, we use the APPLgrid fast NLO interface~\cite{Carli:2010rw} to dynamically calculate the $W$ and $Z$ production cross sections during the fits. For more details, see Ref.~\cite{Accardi:2023gyr}.

\subsection{Deuteron Structure Function}

We evaluate the deuteron DIS structure function $F_{2D}$ in the nuclear impulse approximation \cite{Kulagin:1989mu,Kulagin:1994fz,Kulagin:2004ie,Kulagin:2007ph,Kahn:2008nq,  DelDotto:2016vkh,Fornetti:2023gvf}, namely assuming that the exchanged photon scatters off a bound, off-shell nucleon inside the deuteron. In this approximation, $F_{2D}$ is given by
\begin{equation}
    F_{2D} (x, Q^2) = \sum_{N=p,n} \int d y d p_T^2 \, f_{N/D}(y, p_T^2; \gamma) \, F_{2N} \bigg (\frac{x}{y}, Q^2, p^2 ; \overline\gamma \bigg ),
\label{e:F2D}
\end{equation}
where $x=M_D \, Q^2 / (M \, P_D\cdot q)$ is the \textit{per-nucleon} Bjorken invariant, with $q$ and $P_D$ the photon and deuteron momenta, and $M$ and $M_D$ the nucleon and deuteron masses, respectively. $F_{2N}$ is the structure function of an off-shell nucleon $N$ (a proton $p$ or a neutron $n$) of momentum $p$, $x/y = Q^2/(p\cdot q)$ is its Bjorken invariant, and $y=(p\cdot q)/ (p_D \cdot q)$ is the bound nucleon momentum fraction with respect to the deuteron.
The structure function $F_{2N}$ also depends on the Lorenz-invariant 4-momentum squared $p^2=p \cdot p$ of the bound nucleon, which is off the mass shell with $p^2$ generally smaller than $M^2$.
The integration in the r.h.s. Eq.~\eqref{e:F2D} is obtained under the assumption that final-state interactions are small and the spectator nucleon is on-shell.

The parameter $\overline\gamma = \sqrt{1 + 4 \frac{x^2}{y^2} \frac{p^2}{Q^2}}$ controls the size of target mass effects on the bound-nucleon structure function, and $\gamma = \overline\gamma|_{p^2=M^2,y=1}=\sqrt{1+4x^2 M^2/Q^2}$ determines how the deuteron target's mass affects the bound-nucleon wave function. These factors appear explicitly as arguments of, respectively, the off-shell nucleon structure function and the smearing function $f_{N/D}$. The massless limit is reached as $\gamma,\overline\gamma \to 1$, namely, at momentum transfers $Q^2$ large enough for $x^2 p^2/Q^2 \leq x^2 M^2/Q^2 \to 0$. In that limit target-mass effects become negligible and $f_{N/D}$ can be interpreted as the probability density of finding a bound nucleon $N$ of momentum fraction $y$ inside a deuteron. Away from that limit $\gamma$ is larger than 1, and in DIS kinematics typically takes values in the $1<y \lesssim 1.4$ range~\cite{Accardi:2009br}. 

\subsubsection{Deuteron Smearing}
The smearing function $f_{N/D}$ can be calculated in the Weak Binding Approximation (WBA)~\cite{Kulagin:1989mu,Kulagin:1994fz,Kulagin:2004ie,Kulagin:2007ph,Kahn:2008nq} as a product of the non-relativistic wave function of a nucleon inside the nucleus and kinematic factors that depend on the structure function under consideration. Its full expression is reported in Appendix~\ref{a:nuclWF}.
For our analysis we use a wave function calculated with the AV18 nuclear potential~\cite{Wiringa:1994wb}. As shown Ref.~\cite{Owens:2012bv}, other modern potentials such as WJC-2 \cite{Gross:2010qm} and CD-Bonn \cite{Machleidt:2000ge} produce acceptable, if higher, $\chi^2$ values in a global PDF fit, and the ensuing large-$x$ systematic uncertainty is displayed in Fig.~1 of Ref.~\cite{Li:2023yda}. In this paper we focus on the less studied systematic uncertainty deriving from the implementation of higher-twist and off-shell corrections, and will not further consider the contribution deriving from the choice of nuclear potential. 

The WBA formalism is particularly suitable for describing DIS on weakly bound nuclei such as the deuteron \cite{Owens:2012bv,Ehlers:2014jpa,Accardi:2016qay,Alekhin:2017fpf,Alekhin:2022tip,Accardi:2023gyr} and the tritium, Helium-3 mirror nuclei~\cite{Cocuzza:2021rfn,Alekhin:2022uwc}. It has also been successfully applied to heavier targets \cite{Kulagin:2004ie,Kulagin:2014vsa}.
In general, it is valid if the nucleon that participates in the hard collision is moving with non-relativistic momentum with respect to the nucleus. This happens at $x \lesssim 0.8$, when the bound nucleon itself provides a negligible amount of momentum to the quark that interacts with the virtual photon. At higher values of the Bjorken's invariant\footnote{The per-nucleon Bjorken invariant $x$ in fact ranges between 0 and $\infty$ when $\gamma > 1$, which is always the case. In the strictly massless limit, $\gamma$ $\equiv$ 1, $x$ would range from 0 to $M_D/M$.}, the measurement selects partons of correspondingly higher momentum. Due to the steeply falling nature of the nucleon PDFs, such momentum is increasingly provided by the high-momentum tails of the nucleon wave function instead of by the intrinsic motion of the partons inside the nucleon their are bound to, and eventually a relativistic treatment of the bound nucleon system becomes necessary \cite{Melnitchouk:1994rv,Melnitchouk:1993nk,Gross:2010qm,DelDotto:2016vkh,Fornetti:2023gvf}. Light nuclei maximize the $x$ range where the WBA formalism is applicable due to the weakness of their binding.

\subsubsection{Off-shell corrections}

Deformations of the bound nucleon structure that can affect DIS at large $x$ are assumed to depend on the nucleon's (off-shell) 
four-momentum squared.
In the deuteron, $p^2$ does not differ much from the free-nucleon mass $M^2$ and one can Taylor-expand the off-shell nucleon structure around its on-shell limit $p^2 \simeq M^2$ using the nucleon's virtuality $v=(p^2-M^2)/M^2$ as an expansion parameter~\cite{Kulagin:1994cj,Kulagin:1994fz}. This can be done at the parton level or at the structure function level, namely,
\begin{align}
    \phi(x,Q^2,p^2) &= \phi^{\text{free}}(x,Q^2) \bigg ( 1 + \frac{p^2 - M^2}{M^2} \delta f (x) \bigg ) \, ,
    \label{e:off_pdf}
    \\
     F_{2N}(x,Q^2,p^2) &= F_{2N}^{\text{free}}(x,Q^2) \bigg ( 1 + \frac{p^2 - M^2}{M^2} \delta F (x) \bigg ) \, ,
      \label{e:off_sf}
\end{align}
where $\phi^{\text{free}} = \phi(x,Q^2,M^2)$ and $F_{2N}^\text{free}=F_{2N}(x,Q^2,M^2)$ are free-nucleon PDFs and structure functions respectively, and the $\delta f$ and $\delta F$ ``off-shell functions'' quantify the parton or nucleon deformation when bound in a nucleus. Analytically, they can be written as
\begin{align}
    \delta f (x) &=  \frac{\partial \ln \phi(x,Q^2,p^2)}{\partial \ln p^2} \bigg |_{p^2=M^2} \, ,
    \label{e:deltaf-parton}
    \\
    \delta F (x) &= \frac{\partial \ln F_{2N}(x,Q^2,p^2)}{\partial \ln p^2} \bigg |_{p^2=M^2} \,  .
    \label{e:deltaf-strfn}
\end{align}
In practice $\delta f$ or $\delta F$ are parametrized and fitted to experimental data through the smearing formula \eqref{e:F2D}.
Since the off-shell functions depend on the $\partial \phi/\phi$ or $\partial F_{2N}/F_{2N}$ ratio of PDFs or structure functions, we assume that $Q^2$ evolution effects approximately cancel and we will only fit their $x$ dependence. One can, in fact, heuristically expect that off-shell nucleon deformations originate from low-energy nuclear interactions and be independent of the hard-scattering scale $Q^2$. QCD radiative effects may contribute to the off-shell function at subleading order, but present data are not precise enough to resolve this.
Finally, if only deuteron DIS data are considered among other light nuclei, as in the present global analysis, the fit has no constraints on the flavor dependence of the off-shell functions so that none is indicated in Eqs.~\eqref{e:deltaf-parton}-\eqref{e:deltaf-strfn}. With the inclusion of DIS data on $A=3$ nuclei, the isospin dependence of the off-shell functions could also be studied~\cite{Cocuzza:2021rfn,Alekhin:2022uwc}.

Using the off-shell expansion in Eq.~\eqref{e:off_pdf} or \eqref{e:off_sf} one can obtain a simplified 1-dimensional convolution over just the nucleon's momentum fraction $y$ compared to the 3-dimensional convolution over $y$ and $\vec p_T$ in Eq.~\eqref{e:F2D}. Indeed, by approximating $\overline\gamma \approx \gamma$ in the evaluation of $F_{2N}$ in Eq.~\eqref{e:F2D}, one finds a 1D convolution formula in terms of 2 spectral functions for the on-shell and off-shell components of the nucleon structure function. Schematically, 
$
F_{2D} = \mathcal{S}(\gamma) \otimes F_{2N}(\gamma) 
+ \mathcal{S}^{(1)}(\gamma) \otimes F_{2N}(\gamma) \delta f
$
with details discussed in Appendix~\ref{a:nuclWF}.
This formula considerably speeds up the numerical calculation of the deuteron wave function and has been adopted in the CJ fits, including the present analysis. The error introduced by the approximation of $\overline\gamma$ is of $O(M^2/Q^2)$ and can be absorbed in the fitted higher-twist terms discussed in the next section.    

In the CJ15 and CJ22 analyses, the off-shell nucleon deformations where implemented at the partonic level using a parametrization with two nodes, $\delta f(x) = C(x - x_0)(x - x_1)(1 + x_0 - x)$, as suggested in Ref.~\cite{Kulagin:2004ie}. The parameters $x_0$ and $C$ were fitted but $x_1$ was constrained by imposing the valence quark sum rule at nucleon level. 
However, as in Ref.~\cite{Alekhin:2017fpf}, we noted that we can obtain a better description of the experimental data by adopting a generic polynomial parametrization:
\begin{equation}
    \delta f (x) = \sum_{n=0}^{k} a_{\text{off}}^{(n)} x^n \, ,
    \label{e:off_poly}
\end{equation}
with $a_{\text{off}}^{(n)}$ as free parameters.
In this analysis, we consider a polynomial of second degree ($k=2$, \textit{i.e.}, 3 free parameters) since we found no significant differences in the $\chi^2$ obtained with higher-degree polynomials. The comparison of fitted off-shell functions with $k=2$ and $k=3$ reported in Appendix~\ref{a:poly_comp} rather suggests that experimental data can constrain the off-shell function only in the $x \lesssim 0.7$ region.
This aligns with the region where the Tevatron 
$W$-asymmetry data constrain the baseline free-nucleon 
$d/u$ ratio, which is used in the global fit to determine off-shell deformations in the Deuteron.

We stress that, even in the present flavor-independent implementation, $\delta f$ and $\delta F$ are different beyond LO and should be compared with care. Indeed, the off-shell correction at the structure function level can be written in terms of the one at the partonic level as
\begin{equation}
    \delta F(x,Q^2) \equiv \frac
    {\int_{x}^1 \frac{dz}{z}
        \Big[ \sum_i C_2^i \big( \frac{x}{z} \big) 
         \phi_{i/N}(z,Q^2) \Big] \, \delta f(z) } 
    {\int_{x}^1 \frac{dz}{z}
        \Big[ \sum_i C_2^i \big( \frac{x}{z} \big) \,
        \phi_{i/N}(z,Q^2) \Big] }\ ,
\label{e:OSrel}
\end{equation}
where the sum over $i=q,g$ runs over all parton flavors, and $C_2^i$ are the standard perturbative QCD Wilson coefficients. At LO in the strong coupling constant $\alpha_s$, the Wilson coefficients are proportional to the $\delta(z-1)$ delta function and $\delta F = \delta f$. However, this is no longer true beyond LO, and the full expression \eqref{e:OSrel} can generate a small isospin dependence of $\delta F$ even with a flavor independent $\delta f$. A way to compare off-shell effects determined with different implementations or by different global QCD analyses is discussed in Appendix~\ref{a:off-shell_strfn}.

\subsection{$1/Q^2$ power corrections}
\label{s:1/Q2_corr}

Power corrections to the leading-twist (LT) calculations of DIS structure functions in the large-$x$ region originate from a variety of mechanisms. Most cited among these are target mass corrections (TMCs) and soft parton re-scattering in the final state described by multi-parton matrix elements of twist higher than two. The first ones are of order $O(M^2/Q^2)$ and can be calculated. The second ones are nonperturbative and of $O(\Lambda_{QCD}^2/Q^2)$; they are typically fitted to low-energy DIS data by means of a parametrized ``higher-twist'' term proportional $1/Q^2$. Such phenomenological term will also compensate for other sources of power corrections, as we will review below.

There are several ways to determine TMCs in structure function calculations \cite{Georgi:1976ve,DeRujula:1976baf,Aivazis:1993kh,Kretzer:2002fr,Accardi:2008ne,Steffens:2012jx}, reviewed in Refs.~\cite{Schienbein:2007gr,Brady:2011uy,Ruiz:2023ozv}. When implemented in momentum space, these all involve evaluating the structure functions at the kinematically shifted Nachtmann variable $\xi = 2x_B / (1+\gamma^2)$ instead of at the Bjorken invariant $x_B$. 
The differences among implementations are of order $M^2/Q^2$ and, as shown in Ref.~\cite{Accardi:2009br}, can be effectively fitted in the parametrized higher-twist term alongside the twist-4 multi-parton dynamical effects. 
Power corrections may also arise from the many other choices taken in the setup of the global QCD analysis including, for example, perturbative truncation of the LT calculation \cite{Blumlein:2006be,Blumlein:2008kz,Blumlein:2012se}, phase space limitations at threshold \cite{Corcella:2005us,Accardi:2008ne,Accardi:2014qda,Bonvini:2015ira,Simonelli:2025xpm}, and other phenomenological choices such as the sequential ordering of TMCs and off-shell corrections or the approximations needed to obtain the  1D deuteron convolution formula discussed previously.
For example, missing higher order corrections can be effectively absorbed in the fitted HT terms, that decrease as the perturbative order of the LT calculation increases without substantially affecting the PDF extraction \cite{Martin:2003sk,Blumlein:2006be,Blumlein:2008kz,Blumlein:2012se}.
Therefore, the fitted HT term will not include just the genuine contributions of twist-4 matrix elements to the scattering calculation, and should rather be considered a catch-all function for any \textit{residual power corrections} in the theoretical calculations or their phenomenological implementation in the global QCD analysis framework. The appellative ``higher-twist'' should then be used with care when discussing the phenomenologically fitted power corrections and interpreting their meaning.

We stress that 
as long as a suitable power correction term is additionally included in the fit, the leading twist dynamics can be reliably extracted in terms of PDFs and off-shell functions.
Such extraction should, however, be accompanied by a careful evaluation of the statistical precision of the extracted HT function and its impact on correlated quantities such as the $d/u$ ratio and the quark off-shell deformation $\delta f$. Additionally, a systematic analysis of the (epistemic) uncertainty in the implementation of the HT terms is required -- a task that is inherently complex, as we shall see.

Power corrections are usually implemented in global QCD anlyses by means of \textit{multiplicative} \cite{Virchaux:1991jc,Alekhin:2003qq,Martin:2003sk,Blumlein:2006be,Alekhin:2008ua,Blumlein:2008kz,Accardi:2009br,Owens:2012bv,Blumlein:2012se,Accardi:2016qay,Cocuzza:2021rfn,Accardi:2023gyr} or \textit{additive} \cite{Alekhin:2012ig,Alekhin:2017kpj,Alekhin:2017fpf,Alekhin:2022tip,Alekhin:2022uwc}  modifications of the target-mass-corrected leading-twist structure functions, 
\begin{align}
    F_{2N}^\text{mult}(x,Q^2) &= F_{2N}^\text{TMC}(x,Q^2) \bigg ( 1 + \frac{C(x)}{Q^2} \bigg ) \ , 
    \label{e:ht_mult}
    \\
      F_{2N}^\text{add}(x,Q^2) &= F_{2N}^\text{TMC}(x,Q^2) + \frac{H(x)}{Q^2} \ , 
    \label{e:ht_add}
\end{align}
and fitting the $x$-dependent ``higher-twist functions'' $C(x)$ or $H(x)$ for a nucleon $N=p,n$.  (Variants in which the HT function $C$ multiplies the massless structure functions have also been considered in literature, \textit{e.g.} in Ref.~\cite{Blumlein:2008kz}, with differences of order $M^4/Q^4$ that are not necessarily numerically negligible.)
Note also that fitting the power corrections on top of the calculated TMCs reduces the number of parameters needed for the HT functions \cite{Accardi:2009br}; conversely, TMCs may not be fully captured into a fitted quadratic power-correction term depending on the precision and kinematic reach in $x$ of the experimental data \cite{Krause-thesis:2024}.

It is important to remark that there is no compelling theoretical criterion to choose the multiplicative or additive formula, but any given choice implicitly introduces assumptions on the $Q^2$ scaling and isospin dependence of the HT functions that most often remain unstated. For example, consider an isospin-independent, $Q^2$-independent $C$ coefficient. One can then rewrite Eq.~\eqref{e:ht_mult} in additive terms,
\begin{equation}
    F_{2N}^\text{mult}(x,Q^2) = F_{2N}^\text{TMC}(x,Q^2) + \frac{\tilde{H}_N(x,Q^2)}{Q^2} \ ,
    \label{e:ht_tilde}
\end{equation}
and obtain an equivalent $\tilde H$ additive coefficient which inherits both isospin dependence and $Q^2$ evolution from the $F_2$ structure function, namely, 
\begin{equation}
    \tilde{H}_N(x,Q^2) = F_{2N}(x,Q^2) C(x) \ . 
\label{e:Htilde}
\end{equation}
The same also happens to the equivalent multiplicative term when one assumes an isospin-independent and $Q^2$-independent additive $H$ term. 

As we will discuss in the next subsections, this inherent difference between the treatment of isospin in the multiplicative and additive HT implementations can strongly bias the fit. Further isospin dependence of the HT functions may also arise from other phenomenological implementation choices. For example, in the CJ analyses so far TMCs have been applied for simplicity through the analytic approximation of the Georgi-Politzer formula~\cite{Georgi:1976ve,DeRujula:1976baf} proposed in Ref.~\cite{Schienbein:2007gr}, and the TMC parameter $\overline\gamma$ in the bound nucleon structure function of Eq.~\eqref{e:F2D} is approximated by its on-shell value. Both approximations shift by a slightly different amount the scaling variable and affect differently the proton and neutron structure functions due to their different large-$x$ slopes. The induced residual power corrections may thus also require the use of isospin-dependent HT functions in the global fit~\cite{Alekhin:2003qq,Blumlein:2006be,Alekhin:2008ua,Blumlein:2008kz,Blumlein:2012se,Alekhin:2012ig}.

\subsection{Systematic bias in the implementation of higher-twist corrections}
\label{ss:bias}

Due to the relative scarcity of large-$x$ proton and deuteron data, the HT functions are often assumed to be independent of the isospin of the nucleon participating in the hard scattering. Indeed, previous global studies performed before the publication of JLab DIS measurement indicated only partial sensitivity in the data to isospin-dependent HT effects~\cite{Alekhin:2003qq,Blumlein:2006be,Alekhin:2008ua,Blumlein:2008kz,Blumlein:2012se,Alekhin:2012ig}. 
However, it is difficult to implement isospin-independent power corrections in a self-consistent manner, since many of the effects mentioned in the previous section have different numerical impacts on proton and neutron structure functions. With sufficient sensitivity to isospin coming from Jefferson Lab data~\cite{Cocuzza:2021rfn}, it is now time for a careful re-analysis of power corrections to DIS observables in global QCD fits.

In this section, we discuss in detail the phenomenological necessity of incorporating isospin-dependent HT functions in a global fit. As we will show, failure to do so forces the fit to compensate for the missing isospin-dependent effects by artificially deforming other isospin-sensitive quantities such the deuteron's off-shell function, that only contributes to deuteron observables,  or the $n/p$ or the $d/u$ ratio \cite{Cerutti:2024hrm}. We will analytically study this topic in this section, and in the next one we will numerically verify our result in a global QCD fit. As one can expect, we will see that the deformations will be most prominent at large $x$, where PDFs and structure functions steeply fall to 0. They will also extend to regions well constrained by data, which justifies considering them a systematic implementation bias.

Let us look, in particular, at the neutron-to-proton $n/p$ ratio of the $F_2$ structure functions, whose limiting behavior as $x \to 1$ is sensitive to confinement effects but cannot be directly measured. This ratio may instead be inferred from proton and deuteron target data after removing nuclear corrections: for example, by calculating it in pQCD with the PDFs obtained in the global analysis as we do here, or, similarly, in a data-driven analysis such as in Ref.~\cite{Li:2023yda}. In a global fit, however, both HT and off-shell corrections affect the determination of the deuteron structure function, and biases in either one can be compensated by the fitted parameters of the other, potentially leading to very different determinations of the $n/p$ ratio.

With isospin-independent \textit{multiplicative} HT functions, $C(x) \equiv C_p(x) = C_n(x)$, the correction simply cancels in the $n/p$ ratio and one obtains the same limit as in a LT calculation:
\begin{equation}
    \frac{n}{p} \stackrel{x \rightarrow 1}{ \xrightarrow{\hspace{1cm}} } \frac{4d + u}{4u + d} \simeq \frac{1}{4} 
    \label{e:nop_std}, 
\end{equation}
where we used $d/u \ll 1$ in the $x \to 1$ limit.
For isospin-independent \textit{additive} HT corrections, $H(x) \equiv H_p(x) = H_n(x)$, we obtain instead
\begin{align}
    \frac{n}{p} \stackrel{x \rightarrow 1}{ \xrightarrow{\hspace{1cm}} } \frac{u + 9H/Q^2}{4u + 9H/Q^2} \simeq \frac{1}{4} + \frac{27}{16} \frac{H/u}{Q^2} \ ,
    \label{e:nop_Aiso}
\end{align}
where we used $d/u \ll 1$ again, and in the last step we performed a first-order Taylor expansion in the small but non negligible $H / (uQ^2)$ term. 

Comparing Eq.~\eqref{e:nop_std} to Eq.~\eqref{e:nop_Aiso}, we note that the additive HT produces a larger tail than the multiplicative HT purely due to the phenomenological implementation choice, potentially overestimating the $n/p$ ratio.
Since the neutron $F_{2n}$ structure function only enters in deuteron measurements, this larger tail can be effectively compensated in the fit by a positive off-shell $\delta f$ (or $\delta F$) function\footnote{In principle, a larger neutron $F_{2n}$ tail than needed can also be generated by a larger $d/u$ PDF ratio. This must, however, respect the constraints imposed by other data such as $W$-boson asymmetries in $p+\bar p$ collisions at Tevatron~\cite{D0:2013lql,CDF:2005cgc}, and proton tagged DIS measurements from BONuS\cite{CLAS:2011qvj,CLAS:2014jvt}.}.
Conversely, the multiplicative implementation may lead to an underestimate of $F_{2n}$ that can be compensated by a negative off-shell deformation.
A QCD analysis can then give an equally good description of the included data sets with both HT implementation choices since the current experimental data is insufficient to constrain the off-shell corrections at $x\gtrsim 0.7$, as suggested by the results reported in Appendix~\ref{a:poly_comp}. The price to be paid, however, is the loss of physical interpretability of the large-$x$ off-shell function, and a large systematic uncertainty of the $n/p$ ratio in the same kinematic region.

The discussed HT implementation bias can be removed by considering isospin-dependent HT terms, namely by separately parameterizing $C_p(x)$ and $C_n(x)$, or $H_p(x)$ and $H_n(x)$. Indeed, in the additive implementations one obtains
\begin{align}
    \frac{n}{p} \stackrel{x \rightarrow 1}{ \xrightarrow{\hspace{1cm}} } 
    \frac{u + 9{H}_n/Q^2}{4u + 9{H}_p/Q^2} 
    \simeq \frac{1}{4} + \frac{9}{16} \frac{4{H}_n - {H}_p}{uQ^2} 
    \simeq \frac{1}{4} + \frac{9}{16} \frac{{H}_p/u}{Q^2} \ ,
    \label{e:nop_noiso}
\end{align}
and the same results easily follows by using its equivalent additive representation, namely, by substituting $H$ with $\tilde H$ in the above equation. (In the last step, we estimated $H_n \approx \frac12 H_p$ at large $x$ according to Ref.~\cite{Alekhin:2003qq}.) As a result, the bias in the isospin-independent implementation demonstrated in Eqs.~\eqref{e:nop_std}-\eqref{e:nop_Aiso} is removed, with the $n/p$ tail in Eq.~\eqref{e:nop_noiso} closer to the multiplicative estimate~\eqref{e:nop_std} than to the additive one. 

With the bias removed, we can also expect that the fitted off-shell function will become largely independent of the choice between a multiplicative and additive HT implementation because there is no need of compensation to properly fit the deuteron experimental data. This expectation will be verified in the next section within the CJ22 global QCD analysis framework. 

\section{Interplay of higher-twist and off-shell corrections}
\label{s:results}

In order to test the theoretical expectations discussed in Sec.~\ref{ss:bias}, we have implemented the additive or multiplicative HT scenarios in the CJ22 global analysis framework ~\cite{Accardi:2023gyr}, but using a second order polynomial for better flexibility in the parametrization of the parton-level $\delta f$ off-shell function, see Eq.~\eqref{e:off_poly}. We included the same data sets and we imposed the same kinematic cuts as in the CJ22 analysis, see Ref.~\cite{Accardi:2023gyr} for more details.

Of particular interest for this analysis are the data sets that constrain the large-$x$ $d/u$ PDF ratio and the off-shell deformation $\delta f$ of the nucleons bound in the deuteron. 
The measurements with highest relevance within our global data set are the following: the inclusive DIS structure function for deuteron targets, particularly from SLAC \cite{Whitlow1992}, that is sensitive to both the $d/u$ ratio and to the nuclear corrections; the decay-lepton and the reconstructed $W$-boson charge asymmetries in proton-antiproton collisions at Tevatron \cite{CDF:2005cgc,D0:2014kma,D0:2013xqc,CDF:2009cjw,D0:2013lql}, that are sensitive to the $d/u$ ratio without involving nuclear targets; and the $n/d$ ratio of DIS structure functions from BONuS \cite{CLAS:2011qvj,CLAS:2014jvt}, that probes a nearly free neutron target by measuring the scattered electron in coincidence with a low momentum, backward going spectator proton. The $W$ charge asymmetry can constrain the $d/u$ ratio at $x \lesssim 0.7$ with no need of nuclear corrections
In this region, deuteron DIS data become sensitive to off-shell nucleon corrections. 
The BONuS spectator-tagged measurements, which directly probe the neutron with minimal sensitivity to nuclear corrections, provide an important cross check of the nuclear corrections employed by the fit to calculate the neutron structure function but have limited statistical power by themselves.

With this data in mind, we will first examine the isospin-independent HT implementation, and illustrate the bias discussed in Sect.~\ref{ss:bias}. We will then let the parameters of the proton and neutron HT correction terms vary independently, which will remove the bias. The remaining differences between the additive and multiplicative implementations will then be a measure of the remaining phenomenological systematic uncertainty.

\subsection{Isospin-independent HT implementation}

\begin{figure}[tbh]
\centering
\includegraphics[width=0.97\textwidth]{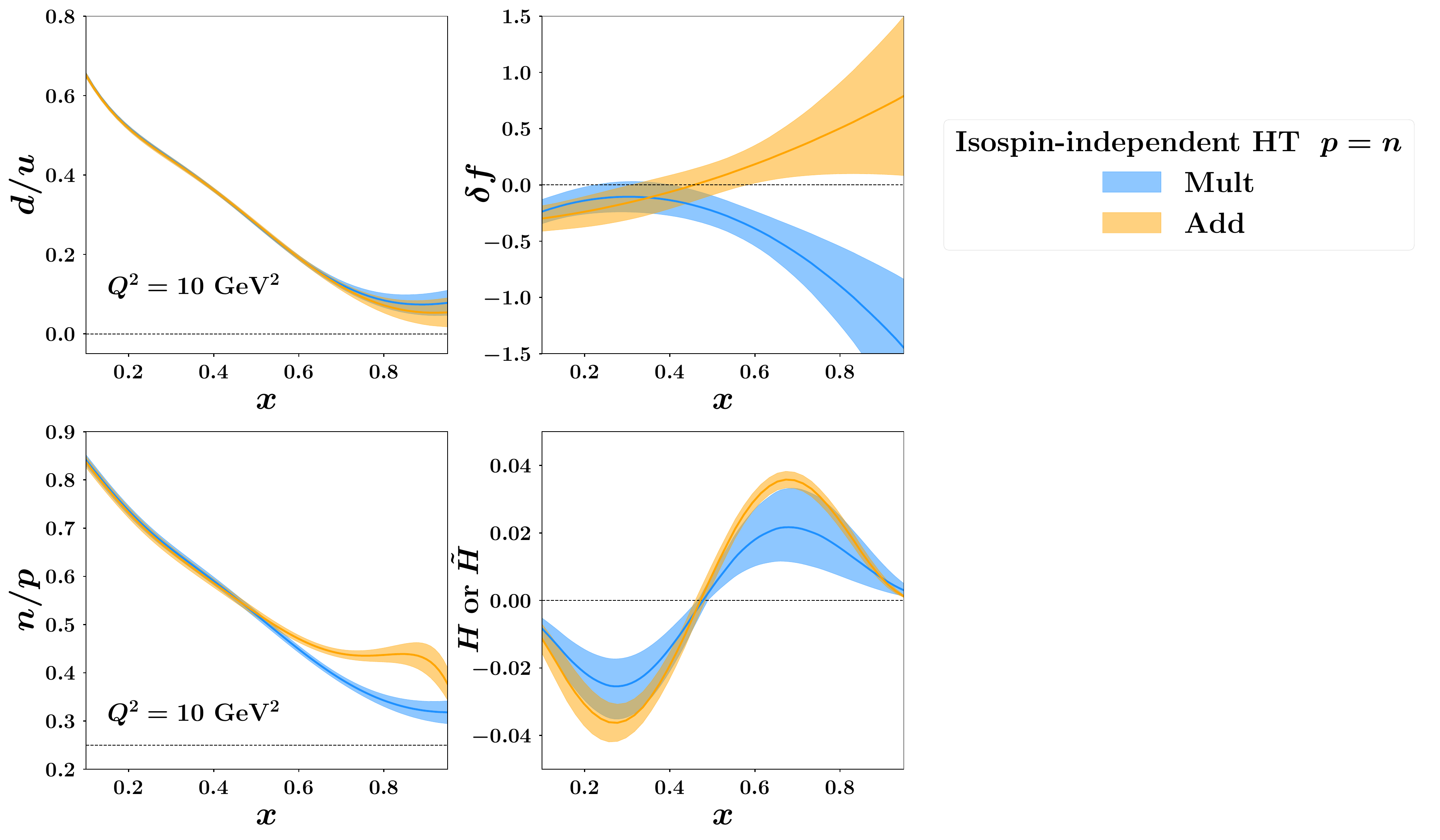}
\vskip-0.2cm
\caption{
Comparison of the results when implementing isospin-independent (HT $p=n$) additive and multiplicative HT corrections, respectively represented by an orange and a blue band. \textit{Upper row}: $d/u$ ratio as a function of $x$ at $Q^2 = 10$ GeV$^2$ (left panel); and off-shell function (right panel). \textit{Lower row}: $n/p$ ratio of $F_2$ structure functions at $Q^2=10$ GeV$^2$ (left panel); and higher-twist corrections (right panel). The bands represent $T^2 = 2.7$ fit uncertainties. 
}\label{f:Fit_iso}
\end{figure}

In Fig.~\ref{f:Fit_iso}, we report the results of the isospin-independent fit. In the left panel of the upper row, we display the $d/u$ PDF ratio as a function of $x$ at $Q^2 = 10$ GeV$^2$, and in the right panel the fitted off-shell function. In the left panel of the lower row, we display the $n/p$ ratio of $F_2$ structure functions at $Q^2=10$ GeV$^2$, while in the right panel we plot the higher-twist term. We depict the additive and multiplicative scenarios with orange and light-blue bands, respectively, that represent a tolerance $T^2 = 2.7$ in the fit's uncertainty~\cite{Owens:2012bv}. We plot the equivalent $\tilde H$ higher-twist function for the multiplicative HT (see Eq.~\eqref{e:Htilde}). Since it inherits an isospin dependence from the $F_2$ structure function, we report the \textit{nucleon} $\tilde{H}=\big( \frac{1}{2}F_{2p} + \frac{1}{2}F_{2n} \big) C(x)$ at $Q^2=10$ GeV$^2$ and we propagate the uncertainty accordingly.

We observe a significant difference at $x\gtrsim 0.6$ between the $n/p$ ratios fitted in the two implementations, with the ratio obtained in the additive HT case approximately $25\%$ larger than the multiplicative one at $x=0.8$. At $x>0.8$ one observes a sudden downturn of the $n/p$ ratio due to the additive HT parametrization that forces $H(x)$ to go to 0 as $x \to 1$ and therefore the $n/p$ ratio to go to 1/4. This happens in a region where the available DIS data is sparse and spans a very limited range in $Q^2$, and therefore do not offer strong constraints to the fit.
At the same time, the HT corrections are largely determined by the proton DIS data and their fitted value show a compatible size in the additive and multiplicative implementations. Likewise, the $d/u$ ratio is dominantly determined by the free-nucleon $W$ asymmetry data and does not show a dependence on the implemented HT model. 

We conclude that the observed pronounced difference in the large-$x$ tail of the $n/p$ ratio is a consequence of the specific phenomenological implementation of the higher twists. And indeed, as expected, the behavior of the $n/p$ ratio is correlated to the size and sign of the extracted off-shell function at $x\gtrsim 0.5$, which is large and positive in the additive scenario while large and negative in the multiplicative one in a region that we have argued is well covered by experimental data. This corroborates our theoretical expectation from 
Sec.~\ref{ss:bias} that the isospin-independent HT implementation produces a bias in the fit results\footnote{A similar difference in the extracted off-shell correction is visible when comparing the CJ (multiplicative) fits~\cite{Accardi:2016qay,Accardi:2023gyr} with the AKP (additive) fits ~\cite{Alekhin:2017fpf,Alekhin:2022tip,Alekhin:2022uwc}. These fits, however,  differ from each other in many respects, making it difficult to pinpoint the source of this discrepancy. Here we compare, instead, the additive and multiplicative implementations within the same fitting framework.}.

\begin{figure}[t]
\centering
\includegraphics[width=0.55\textwidth]{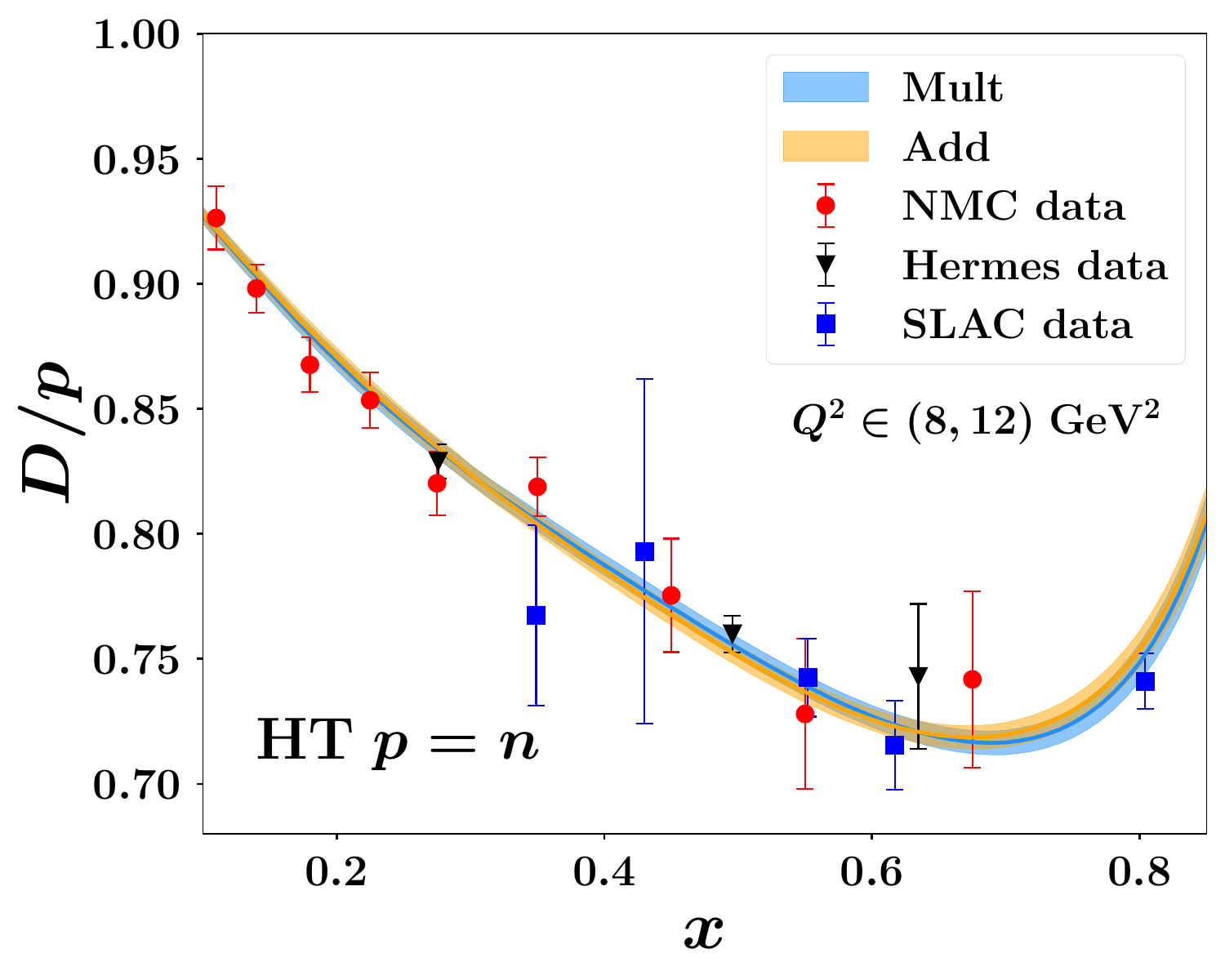}
\vskip-0.2cm
\caption{
Comparison between a selection of experimental measurements of the $D/p$ ratio made by the NMC Collaboration (red points), HERMES Collaboration (black points) and SLAC (blue points), and the theoretical calculations obtained when implementing isospin-independent additive (orange band) or multiplicative (light-blue band) HT corrections (HT $p=n$) in teh global fit. The bands represent $T^2 = 2.7$ uncertainties.}
\label{f:Dop_iso}
\end{figure}

The quantities just discussed are not directly observable, so it is worthwhile looking at how the fits describe relevant scattering processes.
In Fig.~\ref{f:Dop_iso}, we display the comparison between selected data on the $D/p$ ratio of $F_2$ structure functions from the NMC collaboration~\cite{NewMuon:1996fwh,NewMuon:1996uwk}, the HERMES collaboration~\cite{HERMES:2011yno} and SLAC experiments~\cite{Whitlow1992}, and the $D/p$ ratio calculated with the fitted PDFs, off-shell function and isospin-independent HT corrections. 
We can immediately see that the theoretical $D/p$ curves obtained in the additive and multiplicative HT implementations are nearly identical. This result may seem at first incompatible with the discrepant $n/p$ structure function ratio displayed in Fig.~\ref{f:Fit_iso}, where a significantly higher tail was observed at large $x$ for the additive HT implementation.
Instead, it confirms that the larger $n/p$ tail intrinsically induced by the chosen HT implementation can be effectively compensated in the fit of deuteron data by a positive off-shell $\delta f$ (or $\delta F$) function, which gives a negative contribution to $D/p$ and restores the agreement between data and theoretical calculation. Conversely, the multiplicative HT implementation leads to an underestimation of $F_{2N}$ that is compensated by a negative off-shell deformation, providing a description of the experimental data compatible to the additive case. 

\begin{figure}[tbh]
\centering
\includegraphics[width=1.0\textwidth]{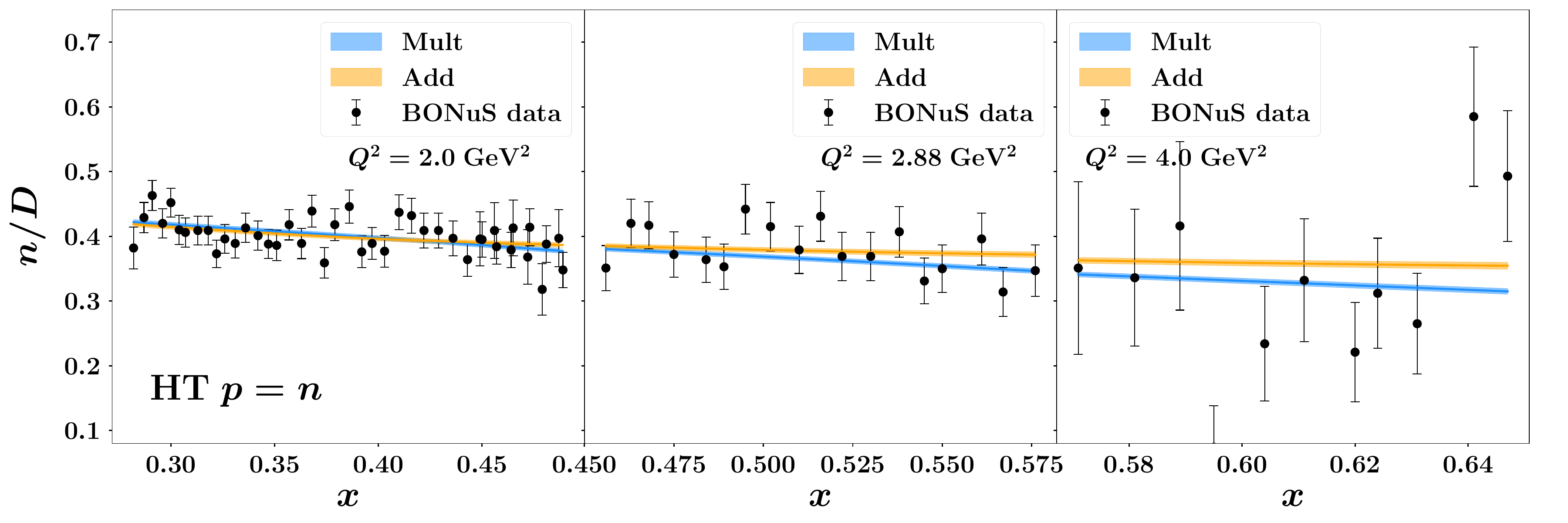}
\vskip-0.2cm
\caption{
Comparison between a selection of experimental measurements of the $n/D$ ratio from the BONuS experiment (black points) and the theoretical calculations obtained when implementing isospin-independent additive (orange band) or multiplicative (light-blue band) HT corrections (HT $p=n$).
The bands represent uncertainties with a $T^2 = 2.7$ tolerance value. 
}
\label{f:noD_iso}
\end{figure}

In Fig.~\ref{f:noD_iso}, we display a selection of experimental data on the $n/D$ ratio of $F_2$ structure functions from the BONuS experiment~\cite{CLAS:2011qvj,CLAS:2014jvt} and compare them with the calculated $n/D$ ratio. The theoretical description is robust, as in the case of the $D/p$ ratio just discussed, but a small and increasing difference can be observed between the additive and multiplicative cases as $x$ increases. This can be expected because the BONuS data are directly sensitive to the $n/p$ ratio. The statistical precision of this data is, however, insufficient to directly discriminate between the multiplicative and additive implementations. For this, new precise data at $x>0.5$ are needed, such as, for example, from the BONuS12 experiment \cite{BONUS12_proposal,Hattawi2024_CLAS}.
%

In summary, these results prove that a large systematic uncertainty on the $n/p$ ratio is introduced in global QCD analyses that assume isospin-independent HT corrections for protons and neutrons. Within this assumption, the choice of an additive or a multiplicative HT implementation significantly impacts the large-$x$ tail of the $n/p$ ratio of structure function. The ensuing large  discrepancy in the obtained ratio is correlated to the large-$x$ behavior of the fitted off-shell function, which is effectively used by the fit to compensate for the over- and under-estimates of the calculated $D/p$ ratio in a kinematical region where there is no direct information on $\delta f$. Should one want to use isospin-independent HT corrections in a fit, it is important to report this source of systematic uncertainty. Nevertheless, we think that a better strategy is to utilize isospin dependent HT terms in the fit to minimize this uncertainty, as we shall discuss next.

\subsection{Isospin-dependent HT implementation}

In Sec.~\ref{ss:bias}, we identified a possible way to reduce the bias on the $n/p$ ratio introduced by the choice of the phenomenological implementation of the HT corrections. Namely, we expect compatible $n/p$ ratios in the additive and multiplicative HT scenarios when these corrections are implemented separately for the proton and neutron structure functions.

\begin{figure}[b]
\centering
\includegraphics[width=0.97\textwidth]{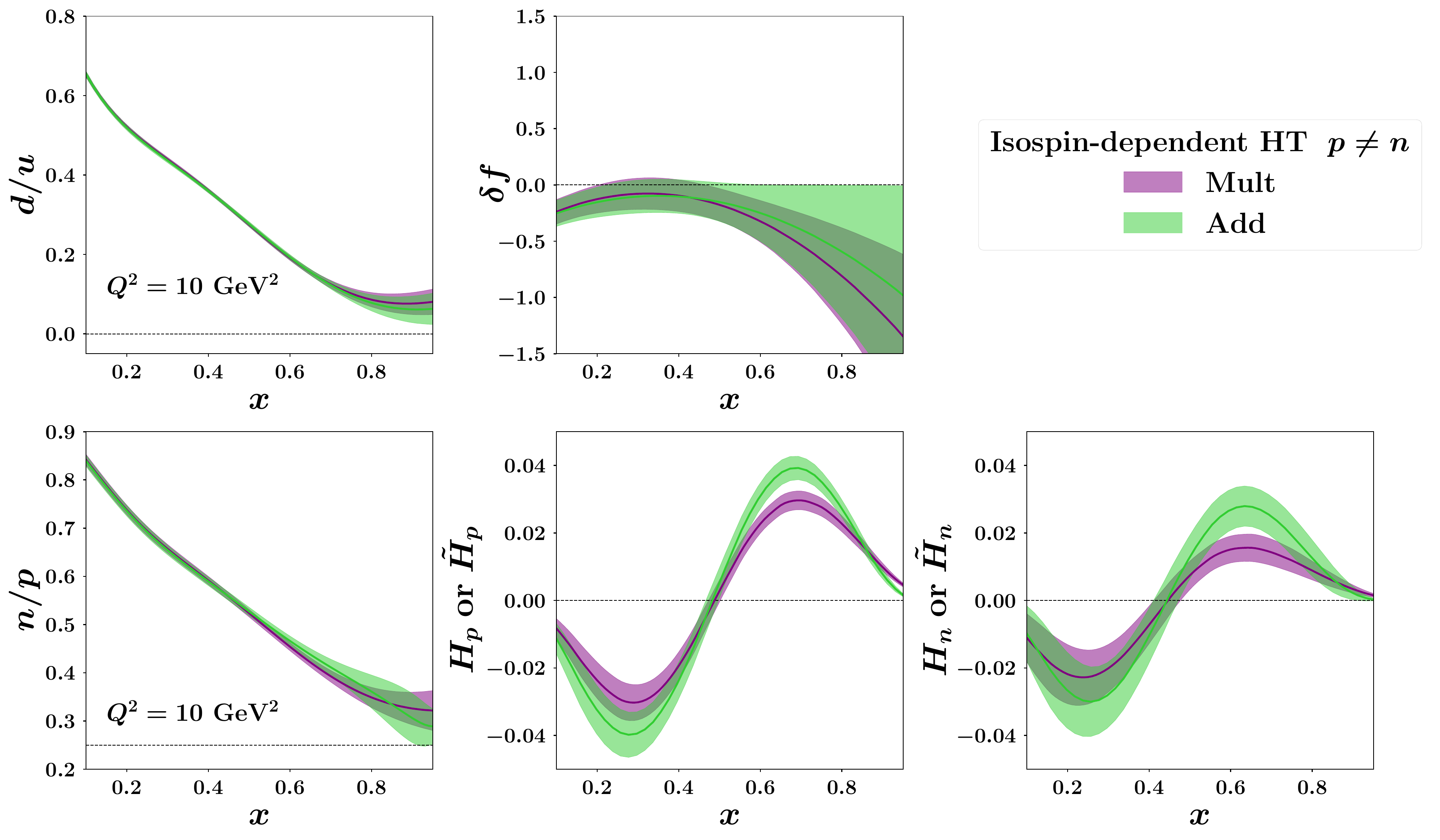}
\vskip-0.2cm
\caption{
Comparison of the results when implementing isospin-dependent (HT $p\neq n$) additive (green band) or multiplicative (violet band) HT corrections. \textit{Upper row}: $d/u$ ratio as a function of $x$ at $Q^2 = 10$ GeV$^2$ (left panel); off-shell function (right panel). \textit{Lower row}: ratio $n/p$ of $F_2$ structure functions at $Q^2=10$ GeV$^2$ (left panel); proton higher-twist correction (central panel); neutron higher-twist correction (right panel).
Bands represent $T^2 = 2.7$ uncertainties.}
\label{f:Fit_noiso}
\end{figure}

In Fig.~\ref{f:Fit_noiso}, we plot the same quantities considered in Fig.~\ref{f:Fit_iso} but extracted from a global fit implementing isospin-dependent HT corrections. The additive and multiplicative HT results are depicted with green and violet bands, respectively, and the proton and neutron HT functions are now plotted in two separate panels. 

We observe that the $n/p$ ratio extracted with additive and multiplicative HTs now agree with each other; they furthermore display a tail which is in between those in Fig.~\ref{f:Fit_iso} but closer to the multiplicative one, as expected from the analysis of Section~\ref{ss:bias}. The extracted off-shell functions are also stable, the correlation with the $n/p$ ratio is significantly reduced and the results in the two HT implementation cases are now compatible with each other.
As explained in Sec.~\ref{ss:bias}, this is a consequence of the isospin-dependent phenomenological implementation of the higher-twist corrections, which provides stability in the estimate of the $n/p$ ratio.

Comparing the lower right and central panels of Fig.~\ref{f:Fit_noiso}, we note that the fitted proton and neutron HT functions are different, with the neutron correction factor about $50\%$ of the proton factor at $x \simeq 0.6$. This is consistent with the assumption used in the last step of Eq.~\eqref{e:nop_noiso}.
The HT terms, now separately fitted for proton and neutrons, absorb the naturally occurring isospin dependent $1/Q^2$ corrections discussed in Section~\ref{s:1/Q2_corr}, and resolve the self consistency issue in the isospin-independent implementation of HT correction discussed in Section~\ref{ss:bias}, namely that an isospin-independent multiplicative correction is isospin-dependent when  represented additively and vice versa. 

Due to the varied nature of the effects contributing to the difference in the proton and neutron HT corrections, we urge caution in interpreting the size and shape of the fitted HT functions. Nonetheless, one can now more confidently utilize the extracted $n/p$ structure function ratio and off-shell functions and quote the small differences obtained in the multiplicative and additive (isospin-dependent) HT implementation as systematic uncertainty. 

\begin{figure}[tbh]
\centering
\includegraphics[width=0.55\textwidth]{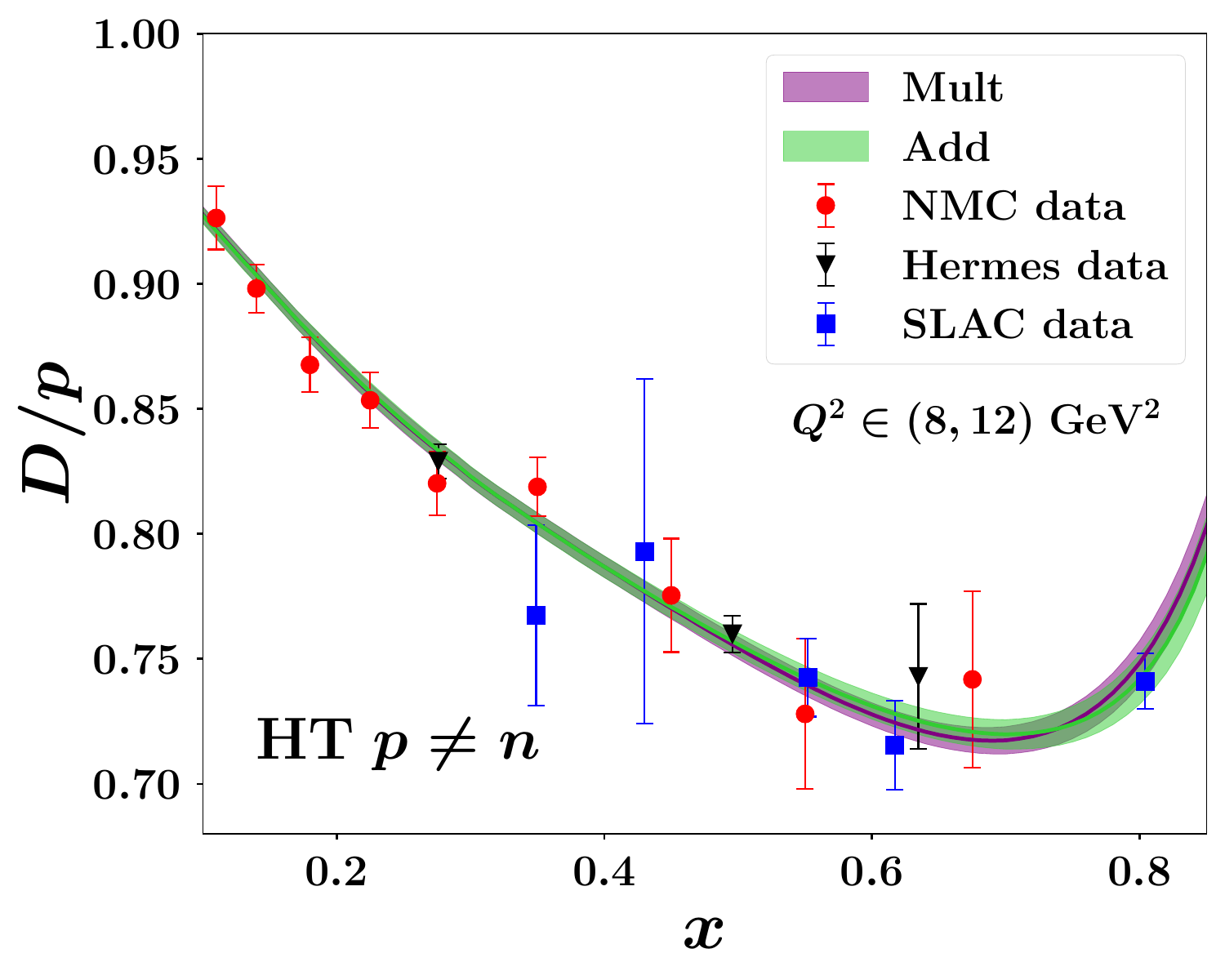}
\vskip-0.2cm
\caption{
Comparison between a selection of experimental data on $D/p$ ratio from the NMC Collaboration (red points), HERMES Collaboration (black points) and SLAC (blue points) and the results of our analyses when implementing isospin-dependent additive (green band) or multiplicative (violet band) HT corrections (HT $p\neq n$).
Bands represent $T^2 = 2.7$ uncertainties.}
\label{f:Dop_noiso}
\end{figure}

\begin{figure}[tbh]
\centering
\includegraphics[width=1.0\textwidth]{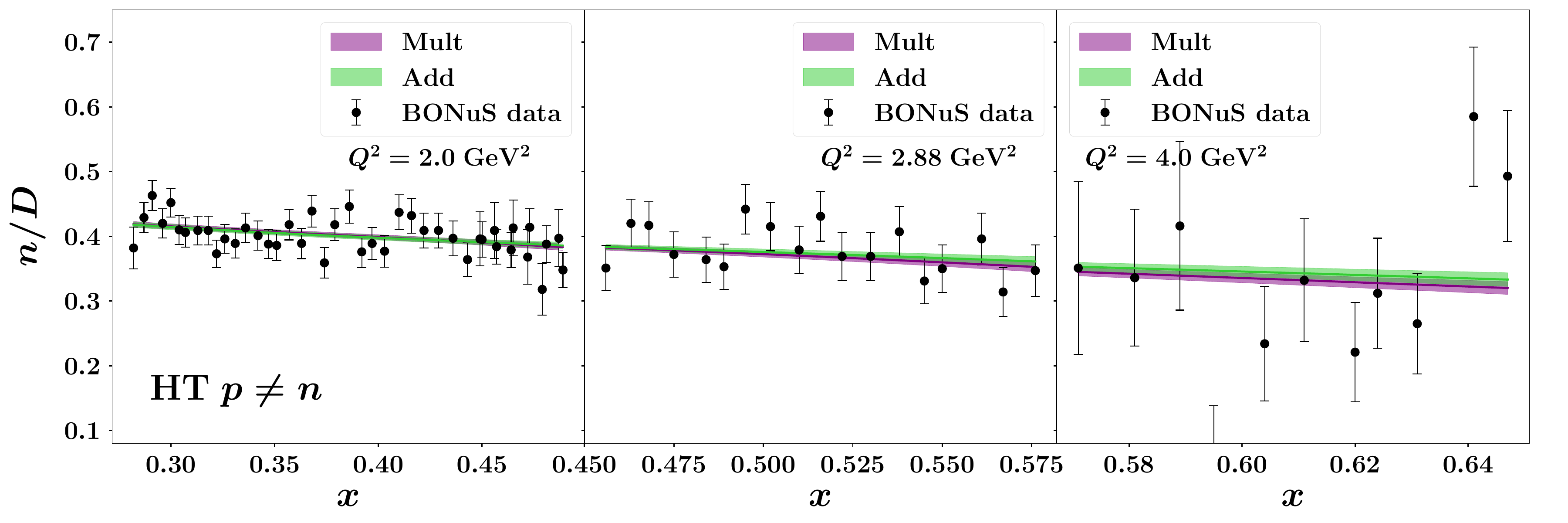}
\vskip-0.2cm
\caption{
Comparison between a selection of experimental data on $n/D$ ratio from the BONuS experiment (black points) and the theoretical results of the these analyses when implementing isospin-dependent additive (green band) or multiplicative (violet band) HT corrections (HT $p\neq n$).
Bands represent $T^2 = 2.7$ uncertainties.}
\label{f:noD_noiso}
\end{figure}

To further support our conclusions, we report in  Figs.~\ref{f:Dop_noiso} and \ref{f:noD_noiso} the $D/p$ and $n/d$ ratios obtained in the isospin-dependent HT fits, and compare these to experimental data. The theoretical calculations are compatible with each other, and their agreement with the experimental data is comparable to that of Figs.~\ref{f:Dop_iso} and \ref{f:noD_iso}, as expected. However, in the isospin-dependent case there there is no need of a different compensation by the off-shell function in the additive and multiplicative implementations to properly describe the experimental data. Furthermore, the $n/p$ structure function ratios obtained in the two implementation are no longer in tension, and compare to the BONuS data in Figure~\ref{f:noD_noiso} in the same manner.

In summary, we have seen that the uncertainties generated by the phenomenological implementation of the HT corrections are significantly decreased when we assume that the HT function of the neutron can be different from that of the proton. At the same time, the agreement between experimental data and theory is preserved, or even improved. The remaining differences between quantities extracted with an additive or multiplicative HT implementation (green and purple bands) can then be used as an estimate of the relatively small systematic implementation uncertainty.

\subsection{Results from other studies}

A brief study of the additive and multiplicative HT implementations was also conducted by AKP in 
Refs.~\cite{Alekhin:2022tip,Alekhin:2022uwc}, but it was limited to the isospin-independent case. 
In contrast to our findings, AKP observed no significant impact of the choice of HT implementation: 
in both cases, their $n/p$ ratio and the structure-function-level $\delta F$ off-shell function exhibit a similar shape 
to our additive fits, shown in orange in Fig.~\ref{f:Fit_iso}.

Although it is challenging to fully analyze the various elements and implementation choices involved 
in global QCD analyses by other groups, we observe that AKP do obtain a statistically significant 
variation in the large-$x$ tail of the $d/u$ PDF ratio when introducing additive or multiplicative 
higher twists (see Figure 4 of Ref.~\cite{Alekhin:2022uwc}). This variation occurs at $x > 0.3$, where 
the statistical power of the decay lepton asymmetry from Drell-Yan process decreases~\cite{Accardi:2016qay}, 
allowing the fit to compensate for the HT bias by a deformation of the $d$-quark PDF. In contrast, in the CJ framework we include also experimental data on kinematically reconstructed $W$-boson asymmetries, which provide strong constraints 
on the $d$-quark distribution up to $x \sim 0.7$. As a result, in our fits, the compensation of the HT bias is shifted 
mostly to the off-shell function.

The off-shell deformation we have fitted is largely compatible with zero. Does this mean that quark densities are not modified by nucleon offshell effects? Not really, because a non-zero offshell deformation of the $u$ quark maybe be cancelled by an opposite deformation of the $d$ quark, as shown in a recent JAM analysis \cite{Cocuzza:2021rfn}. In order to perform a similar flavor decomposition of the $\delta f$ function, however, we would also need to include in the fit the \textsc{MARATHON} experimental DIS data on $^3H$ and $^3He$ targets from Jefferson Lab~\cite{JeffersonLabHallATritium:2021usd}, which we leave for future work.

Finally, it is worthwhile noticing that the off-shell corrections have been assumed to be independent of $Q^2$ in our work as well as in Refs.~\cite{Alekhin:2022tip,Alekhin:2022uwc,Cocuzza:2021rfn}. It is then somewhat peculiar that a change in these (or in the $d/u$ ratio that is only logarithmically dependent on $Q^2$) may compensate for HT corrections that are power suppressed as $1/Q^2$. This is likely due to the fact that at large $x\gtrsim 0.4$, where this compensation takes place, the only DIS data on deuteron target essentially come from SLAC experiments, with a limited range in photon virtuality and limited statistics at large $x$. With higher precision data, such as from Jefferson Lab at 12 GeV \cite{Biswas:2024diw,Dudek:2012vr} or even its envisioned 22 GeV upgrade \cite{Accardi:2023chb}, it should be possible to disentangle the interplay of off-shell corrections and HT model implementations in global fits \cite{Cerutti-talk-Frascati-22GeV}.

\section{Fits excluding $W$-boson asymmetry data}
\label{s:noW}

The suitability of including both the lepton asymmetry and reconstructed $W$-boson asymmetry data in the same fit, as we do in this paper, has been questioned in the literature \cite{Alekhin:2017fpf}: from a statistical point of view as a source of double counting, and from a QCD analysis point of view as possible driver of the differences between the extracted off-shell function presented in the previous section and the results of the AKP analysis. We disagree on both accounts. 

Regarding the first concern, even if it is true that the two experimental analyses have used the same set of recorded events, the $W$-reconstruction procedure also leverages the missing energy of each event to statistically estimate the neutrino's energy and, thus, includes more physical information in the obtained data than the lepton asymmetry analysis. As a consequence, the $W$-asymmetry data are sensitive to a different and complementary $x$ range. Of interest for the present work, $W$ measurements reach higher values $x\lesssim 0.7$ than the $x\lesssim 0.3$ probed by the lepton measurements. A small statistical double counting does exist at smaller $x \approx 0.2$, but it is negligible compared to the gains obtained by using the $W$-asymmetry data set. We will discuss this more in Appendix~\ref{a:doubleC}.

Here, we address in detail the second concern, namely that $W$-asymmetry data may drive the correlation between $d/u$, the HT corrections and the off-shell corrections, requiring the latter to be concave (small and negative) after the HT implementation bias is removed. In practice, we repeat the fits discussed in Section~\ref{s:results} excluding the $W$ asymmetry data. 

\begin{figure}[tbh]
\centering
\includegraphics[width=0.97\textwidth]{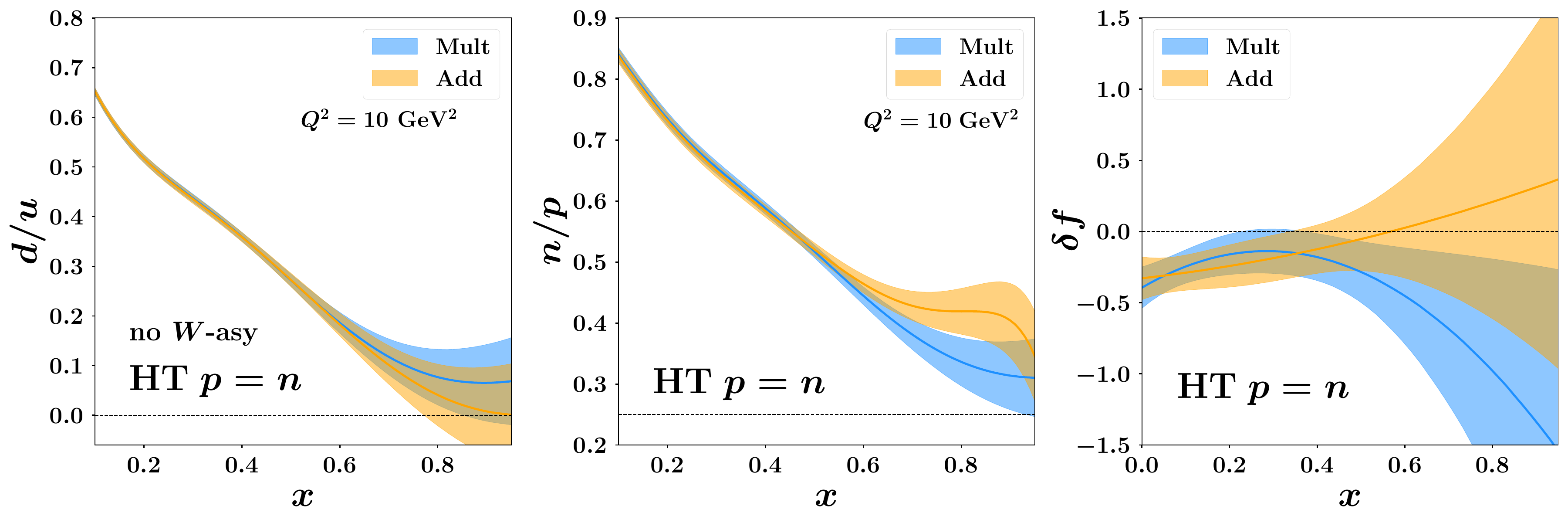}
\vskip-0.2cm
\caption{
Comparison of the results of the CJ analyses when excluding $W$-boson asymmetry experimental data and implementing isospin-independent additive (orange band) or multiplicative (light-blue band) HT corrections (HT $p=n$). Left panel: $d/u$ ratio as a function of $x$ at $Q^2 = 10$ GeV$^2$; central panel: ratio $n/p$ of $F_2$ structure functions at $Q^2=10$ GeV$^2$; right panel: off-shell function.
Bands represent $T^2 = 2.7$ uncertainties.
}
\label{f:Fit_noW}
\end{figure}

In Fig.~\ref{f:Fit_noW}, we report the results for the isospin-independent HT correction fits (HT $p=n$). We focus on the quantities of interest, that is the $d/u$ PDF ratio (left panel), the $n/p$ structure function ratio (central panel), and the off-shell function (right panel). The fits with multiplicative HT corrections are displayed in blue, and in orange we show the additive correction fits.
The result is very similar to that presented in Fig.~\ref{f:Fit_iso}, but with larger uncertainty bands on the extracted quantities. This is particularly visible at $x\gtrsim 0.4$ where the fit no longer can utilize the constraints provided by the $W$ boson data. The off-shell corrections become compatible with each other within the fit uncertainty even though their central values has scarcely changed. However, the $n/p$ ratios remain well separated at large $x$, showing that the difference in the fitted HT corrections is not data driven but rather intrinsic to the phenomenological implementation choice, as we argued from a general standpoint in Section~\ref{ss:bias}. 

\begin{figure}[tbh]
\centering
\includegraphics[width=0.97\textwidth]{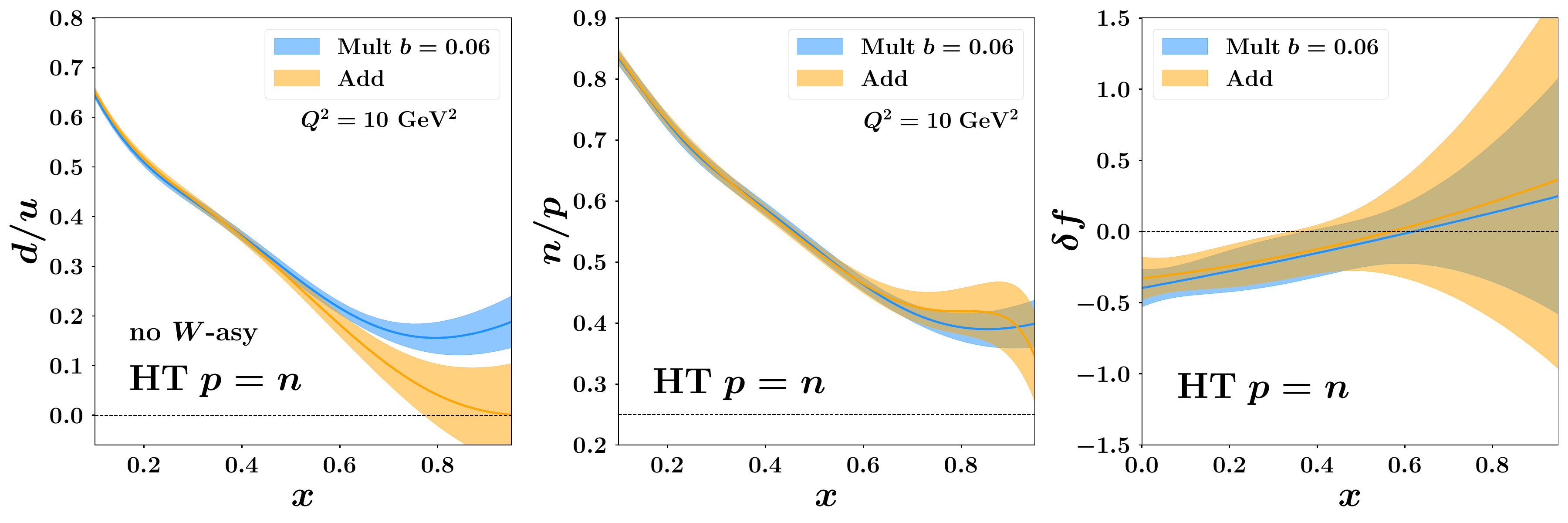}
\vskip-0.2cm
\caption{
Same conventions and notation as in previous figure but imposing $b=0.06$ in the parametrization of the $d/u$ tail in the multiplicative HT scenario.}
\label{f:Fit_noW_bfix}
\end{figure}

In order to understand what data drive this result, we performed fits around a configuration that produced an $n/p$ ratio similar to that obtained in the AKP analysis. This can be achieved, for example, by prescribing an artificially high value for the parameter $b$ in the $d$-quark parametrization~\eqref{e:PDF_du}. This is the parameter that directly controls the end-point value of $d/u$ ratio.
The fit results are shown in Fig.~\ref{f:Fit_noW_bfix} for fixed $b=0.06$, that qualitatively reproduces the AKP result (except for the end-point limit $d/u=0$ at $x=1$). The multiplicative $n/p$ ratio and $\delta f$ off-shell function increase compared to our nominal fit and become compatible with their additive fit counterparts. But the HT implementation bias now shows up in the $d/u$ panel, as it happens in the fits by AKP. Yet, in our case, this inflated-$b$ fit is disfavored by the DIS deuteron data that drive a $\Delta\chi^2 =10$ change in the total chi squared of the fit. The $W$-asymmetry data, are in even larger disagreement with this fit, with a calculated $\chi^2_W/\text{npt} = 6$. 

A similar result can also be obtained by artificially increasing the multiplicative off-shell function to make it compatible with the additive one, but letting the $b$ parameter free: the multiplicative $n/p$ is compatible with its additive counterpart, but the $d/u$ ratio remains different from the additive $d/u$, even with a slightly larger uncertainty than in Fig.~\ref{f:Fit_noW_bfix}. As in the previous fit, we observed a $\Delta\chi^2 = 10$ increase driven by DIS deuteron data, and calculated $\chi^2_W/\text{npt}=6$ for the $W$ asymmetry data that were not included in the fit. 

These studies show that, in restricted sectors of the parameter space, it is possible to reconcile the $n/p$ ratios and $\delta f$ off-shell function obtained in the multiplicative and additive HT $p=n$ fits. Even then, these two HT implementation choices are not equivalent because the DIS deuteron data now force the $d/u$ ratio instead of the off-shell function to compensate for the biases introduced by the isospin independence assumption for HT corrections. The pull that the DIS data exert seems small if one looks at the mentioned $\Delta\chi^2$ values, but it is non-negligible. This is demonstrated by the fits obtained without parameter constraints already presented in Fig.~\ref{f:Fit_noW}: with a smaller $\chi^2$, the DIS data is compatible with the rest of the data sets and the HT bias shows up exactly where one expects it theoretically.

\begin{figure}[tbh]
\centering
\includegraphics[width=0.97\textwidth]{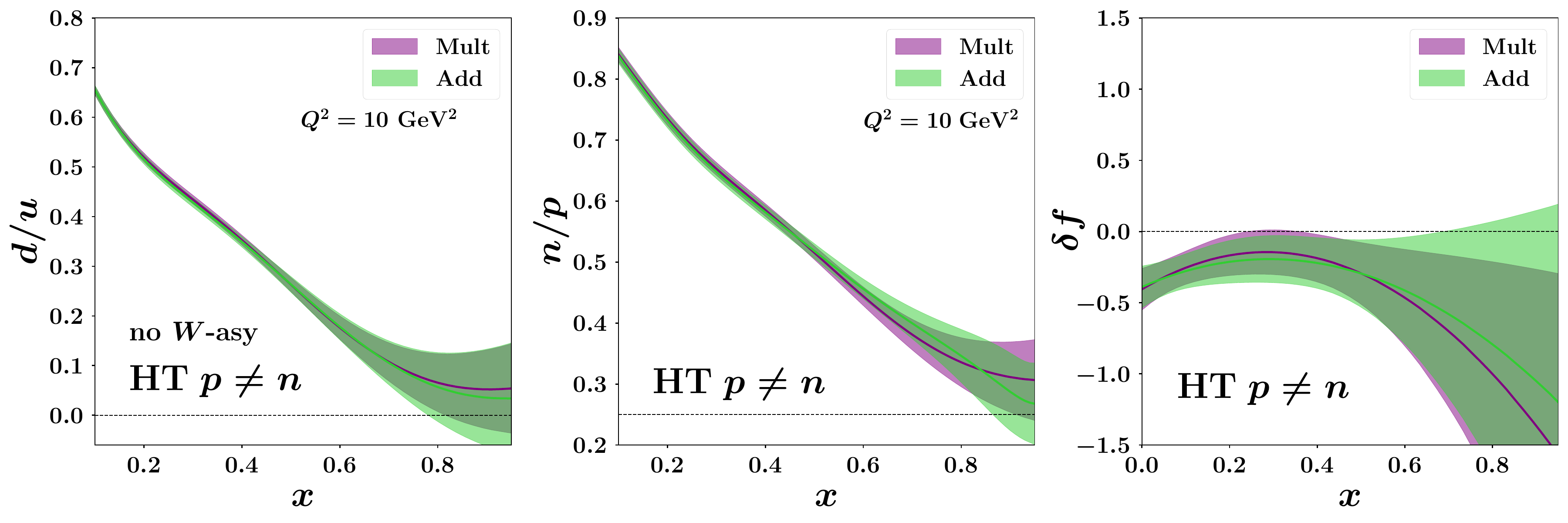}
\vskip-0.2cm
\caption{
Fit excluding $W$-boson asymmetry data and implementing isospin-dependent (HT $p\neq n$) corrections. Same conventions and notation as in Fig.~\ref{f:Fit_noW}.}
\label{f:Fit_noW_noiso}
\end{figure}

We finally confirm that, even when one excludes the $W$-asymmetry data from the analysis, the effects of the bias can only be significantly reduced if the HT corrections are fitted independently for the proton and the neutron. In Fig.~\ref{f:Fit_noW_noiso}, one can indeed see that all 3 relevant quantities (the $d/u$ ratio, the $n/p$ ratio and the off-shell $\delta f$ function) are compatible with each other in both the additive and multiplicative HT fits. They are also compatible with the analysis reported in Fig.~\ref{f:Fit_noiso} that included the $W$-asymmetry data, but with larger uncertainties starting at $x \approx 0.4$.
Increasing the precision of the fit at larger $x$, which is the ultimate goal of the CJ collaboration, thus requires one to also include the $W$-asymmetry data. As discussed in detail Appendix~\ref{a:doubleC}, this is both statistically meaningful and desirable.

\section{Conclusions}
\label{s:conclusion}

In this paper we presented results for four different treatments of the higher-twist corrections to DIS structure functions: additive vs. multiplicative and isospin-independent vs. isospin-dependent. 

We have shown that the behavior of the $n/p$ structure function ratio at large $x$ strongly depends on the choices made for these corrections. In particular, we found that the isospin-independent additive choice leads to a significant increase of the $n/p$ ratio in the large-$x$ region as compared to the isospin-independent multiplicative choice. In the fits to the deuteron data the introduction of parametrized off-shell corrections artificially compensates these differences in the $n/p$ ratio such that good agreement to the $D/p$ data can be obtained for either choice.

By contrast, when the isospin-independence constraint on the HT corrections is relaxed, we find that comparable results for the $n/p$ ratio and the off-shell functions can be obtained using either the multiplicative or additive form for the HT corrections, as shown in Fig.~\ref{f:Fit_noiso}. The result is a small but non zero $d/u$ limit as $x \to 1$, and an off-shell function compatible with 0. With the discussed HT implementation bias so minimized, one can use the differences in the quantities obtained with an additive or multiplicative implementation of the HT corrections to confidently quantify the remaining systematic uncertainty.

We also found that including in the fit the reconstructed $W$ asymmetry data from Tevatron free proton-antiproton collisions is essential to maximize the precision of the $d/u$ ratio extraction at large $x$. This data set is in practice the only direct experimental constraint on $d/u$ from free proton targets at $x\gtrsim 0.4$, since BONuS data lack sufficient statistical power for an effective flavor separation at larger values of $x$. 

The off-shell functions used here have been assumed to be independent of $Q^2$ whereas the HT terms decrease as 1/$Q^2$. Therefore, the correlation between the HT and off-shell corrections discussed in this paper in principle varies with $Q^2$, but is not detectable with the limited statistics and range in photon virtuality of the currently available deuteron DIS measurement. Additional data in the large-$x$ region at higher $Q^2$, or higher precision data in the currently explored range, will allow for further constraints to be placed on the off-shell corrections, and to disentangle these from HT corrections.

The new experimental data on the $D/p$ ratio from Hall C \cite{Biswas:2024diw} and the upcoming $n/D$ ratio from BONuS12 \cite{BONUS12_proposal,Hattawi2024_CLAS}, among other DIS measurements planned with the 12 GeV beam at Jefferson Lab, will be essential to verify our conclusions and to control the interplay of off-shell and HT corrections in global fits. Tagged DIS cross sections differential in the nucleon virtuality $p^2$ could furthermore provide invaluable information on the behavior of off-shell nucleons. Data from the envisioned JLab 22 GeV upgrade will further enhance the precision and kinematic range in $Q^2$ available to large-$x$ global analyses, that will then be able to disentangle the leading-twist offshell effects from $1/Q^2$ power suppressed HT corrections, and shed light on quark and gluon dynamics in nuclei.

Future global analyses that combine these low-energy, high-statistics data with higher-energy data from the EIC \cite{AbdulKhalek:2021gbh} and the fixed-target program of the LHCb experiment \cite{LHCb-fixed-target}, which will be minimally affected by HT corrections without reaching mass scales sensitive to new physics, will then be able to precisely and accurately delineate large-$x$ quark and anti-quark PDFs. 
This will in turn enable both an improved extraction of the HT from the already analyzed lower-energy data and a robust identification of new physics signals, \textit{e.g.}, from large mass and forward particle production measurements at the LHC \cite{Ball:2022qtp,Hammou:2023heg,Hammou:2024xuj}. 


\begin{acknowledgments}
We gratefully aknowledge S.~Alekhin, S.~Kulagin, W.~Melnitchouk and R.~Petti for their collaboration on a benchmark exercise of our codes, which offered non-trivial insights for addressing the topic of this paper. We are also thankful to P. Risse for his thoughtful feedback on the manuscript. 
This work was supported in part by the  U.S. Department of Energy (DOE) contract DE-AC05-06OR23177, under which Jefferson Science Associates LLC manages and operates Jefferson Lab, and by DOE contract DE-SC0025004, DE-AC02-05CH11231.
\end{acknowledgments}


\bibliography{HTvsOSbiblio}
\bibliographystyle{myrevtex}

\newpage
\appendix

\section{Nuclear smearing functions}
\label{a:nuclWF}
Nuclear smearing functions in the WBA approximation are discussed in detail in Ref.~\cite{Kulagin:2004ie}. In the deuteron's rest frame,  the smearing function $f_{N/D}$ used to build the $F_{2D}$ structure function (see Eq.~\eqref{e:F2D}) reads
\begin{align}
  f(y_D,p_T^2,\gamma) & = 
    \theta(y_D) \, \theta(y_D^\text{max} - y_D) \times 
    \theta(p_v^\text{cut} - p_v) 
    \nonumber \\
  & \quad \times \left[\frac14\  \frac{\gamma M_D E_s}{(1-y_D)M_D +(\gamma^2-1) E_s }\right]
    \times  \bigg[ 1+ \frac{\gamma p_z}{M} \bigg] 
    \times\, |\varphi(p_v)|^2
    \nonumber \\
  & \quad \times \frac{1}{\gamma^2}\,
    \Bigg[ 1+ \frac{(\gamma^2-1)}{\left(y_D \, M_D/M   \right)^2}\left( \left(1 + \frac{\epsilon}{M}\right)^2 + \frac{p_v^2 - 3 p_z^2}{2M^2}\right) \!\Bigg] \ .
\label{e:f_WBAREL}
\end{align} 
The function $f_{N/D}$ depends on the nucleon's invariant momentum fraction $y_D =  \frac{p \cdot q}{P_D \cdot q}$, and the nucleon's transverse momentum squared $p_T^2$. The parameter $\gamma = \sqrt{1+4 x^2 M^2/Q^2}$ controls the deuteron mass corrections. The masses $M$ and $M_D$ are, respectively, the free nucleon's and the deuteron's. At the right hand side $\epsilon = 2.23$ MeV is the deuteron's binding energy,
\begin{align*}
  p_v 
    & = \sqrt{p_T^2+p_z^2} \\
  p_z
    & = \frac{1}{(\gamma^2-1)}
        \Big[ -M_D(1-y_D)\gamma + \sqrt{(1-y_D)^2M_D^2 + (\gamma^2-1)(p_T^2+M^2)} \Big] 
\end{align*}
and
\begin{align*}
  y_D^{max} & = 1 \\
  p_v^{\text{max}} & = 1200\ \text{MeV}  
\end{align*}  
are numerical phase space cutoffs ($y_D$ is in fact unlimited, and $p_v$ is limited only if non-relativistic kinematics is used).
Assuming negligible final state interactions, the spectator nucleon is on shell with energy
\begin{align}
     E_s & = \sqrt{p_v^2+M^2} \ .
\end{align}
The nucleon wave function $\varphi$ is normalized such that $1 = \int dp_v p_v^2 |\varphi(p_v)|^2$. This normalization absorbs the angular integration in the definition of $\varphi$. (Another common normalization, not used here, is $1=\int d^3p |\psi|^2$, where $\psi = \varphi / \sqrt{4\pi}$.) 
Baryon number conservation is imposed by multiplying the smearing function by $M/p_0$, where $p_0 = \sqrt{M_D^2 - E_s^2}$ is the energy of the active, off-shell nucleon.

The smearing function is most easily written in terms of to the unscaled deuteron kinematics. It is however customary, as we did in Eq.~\eqref{e:F2D}, to express it using the per-nucleon momentum fraction $y = (M_D/M)\, y_D$. The per-nucleon smearing function of Eq.~\eqref{e:F2D} is then defined by $f_{N/D}(y,p_T^2) \, dy = f_{N/D}(y_D,p_T^2) \, dy_D$, and reads
\begin{equation}
    f_{N/D}(y) = \frac{M}{M_D} f_{N/D}(y_D) \ .
\label{e:CJ_vs_AKP}
\end{equation}

Upon the off-shell expansion \eqref{e:off_pdf} or \eqref{e:off_sf}, one can perform the $dp_T^2$ integrals in the convolution formula \eqref{e:F2D} and obtain a 1D convolution:
\begin{align}
    F_{2D}(x,Q^2) = \int dy \mathcal{S}(y,\gamma) \otimes F_{2N}\big({\textstyle \frac{x}{y}},Q^2;\gamma \big)
        + \int dy \mathcal{S}^{(1)}(y) \otimes F_{2N}\big({\textstyle \frac{x}{y}},Q^2,\gamma \big)       
       \delta f\big({\textstyle \frac{x}{y}} \big) \ ,
\label{eq:smearing_off}
\end{align}
where
$
    \mathcal{S}(y;\gamma) \equiv \int dp_T^2 \, f_{N/D}(y_D,p_T^2;\gamma) 
$
and
$
    \mathcal{S}^{(1)}(y;\gamma) \equiv \int dp_T^2 \, \frac{p^2-M^2}{M^2} \, f_{N/D}(y,p_T^2;\gamma) 
$
are $y$-dependent spectral functions, respectively, for the on-shell and off-shell components of the nucleon wave-function.

\section{Third order polynomial parametrization for the off-shell function}
\label{a:poly_comp}

In this Appendix, we report the results for the extraction of the off-shell function in the various scenarios discussed in the paper. In particular, we focus on the differences obtained when the off-shell is parametrized as a polynomial function of 2$^{nd}$ or 3$^{rd}$ degree (see Eq.~\eqref{e:off_poly}).

\begin{figure}[tbh]
\centering
\includegraphics[width=0.97\textwidth]{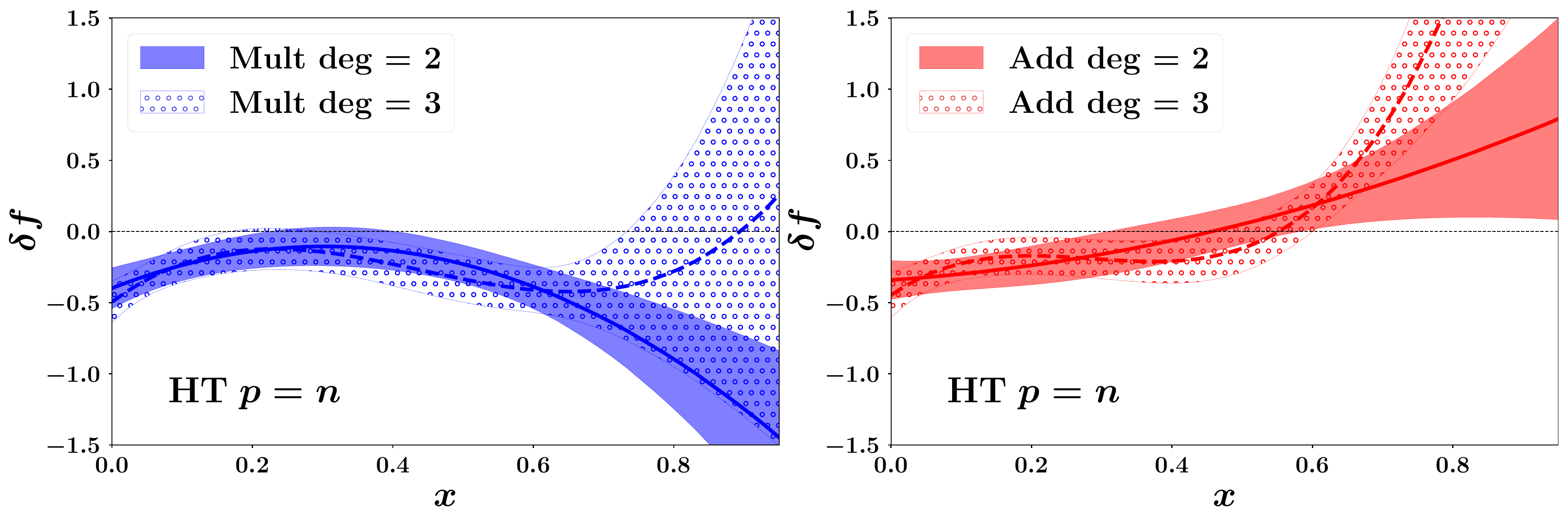}
\vskip-0.2cm
\caption{
Comparison of the extracted off-shell function with polynomial parametrization of 2$^{nd}$ (plain band) and 3$^{rd}$ degree (dotted band) when implementing isospin-independent (HT $p = n$) HT corrections. \textit{Left panel}: multiplicative HT scenario. \textit{Right panel}: additive HT scenario.
Bands represent $T^2 = 2.7$ uncertainties.}
\label{f:poly_iso}
\end{figure}

In Fig.~\ref{f:poly_iso}, we show the extracted off-shell function $\delta f$ in the isospin-independent case for HT corrections (HT $n = p$). The baseline result obtained with the polynomial of 2$^{nd}$ degree is depicted with a plain blue (red) band for multiplicative (additive) HT implementation, while with a dotted band for the polynomial of 3$^{rd}$ degree. We note that the two bands in both the left and right panels are mostly compatible with each other and start to deviate at large $x$. Since the total $\chi^2$ of the two cases are almost equivalent, we interpret this as evidence that the off-shell function is unconstrained beyond $x \simeq 0.7$. In this region, in fact $W$-boson data no longer constrain the free-nucleon $d/u$ ratio that the fit leverages to extract the off-shell function from deuteron DIS data.

\begin{figure}[tbh]
\centering
\includegraphics[width=0.97\textwidth]{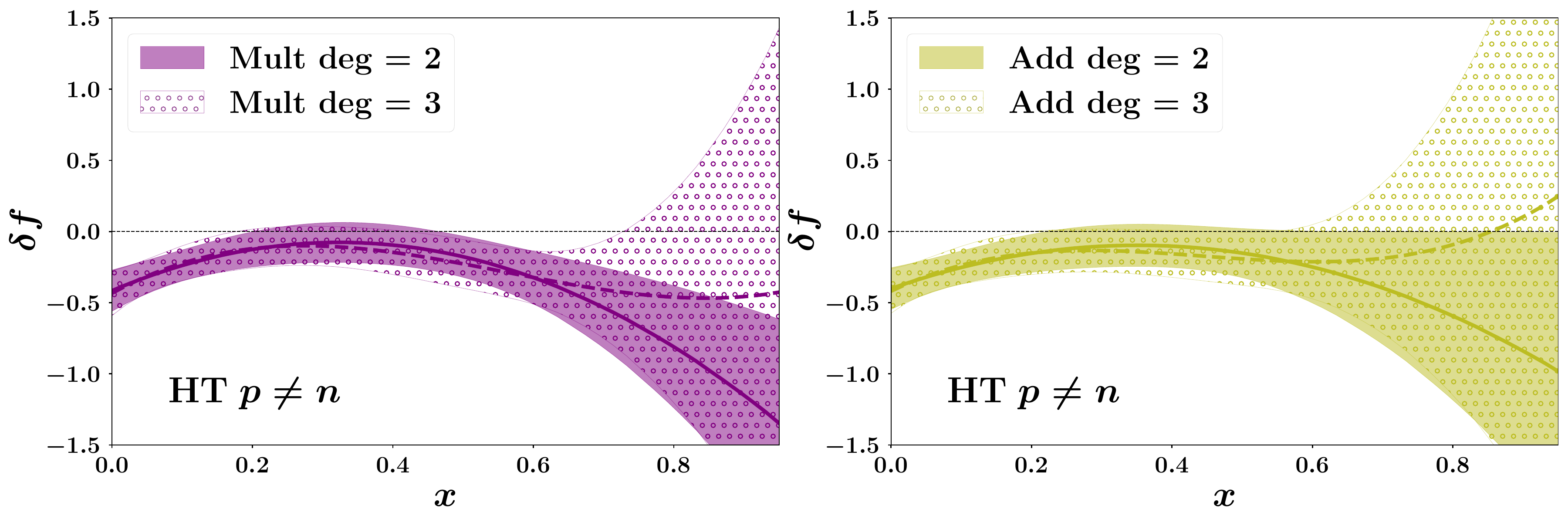}
\vskip-0.2cm
\caption{
Same as Figure~\ref{f:poly_iso} but isospin-dependent (HT $p \neq n$) HT corrections.}
\label{f:poly_noiso}
\end{figure}

In Fig.~\ref{f:poly_noiso}, we show the extracted off-shell function $\delta f$ in the isospin-dependent case for HT corrections (HT $n \neq p$). The baseline result obtained with the polynomial of 2$^{nd}$ degree is depicted with a plain purple (green) band for multiplicative (additive) HT implementation, while with a dotted band for the polynomial of 3$^{rd}$ degree. Once again, we note that the two bands in both the left and right panels are compatible with each other and start to slightly deviate beyond $x \simeq 0.7$.

Since there are no significant differences between the off-shell functions extracted with second and third degree polynomial parametrizations in the region covered by currently available experimental data, we choose to use the polynomial function of 2$^{nd}$ degree as a baseline for the study discussed in the paper. This choice is also supported by the fact that the $\chi^2$ value is not improved by the introduction of the additional parameters needed for the 3$^{rd}$-degree polynomial function.

\section{Off-shell corrections at structure function level}
\label{a:off-shell_strfn}

As discussed in the main text, the off-shell function $\delta f$ or $\delta F$ are not directly observable and, furthermore, they correlate strongly with the $d$-quark PDF and with the higher-twist corrections. Therefore, a direct comparison of results obtained in different global QCD analyses is not straightforward.

A more straightforward comparison may be obtained with an ``effective'' off-shell correction to the deuteron $F_{2D}$ structure function dwefined as  \textit{i.e.}
\begin{equation}
\delta F_{2D} (x,Q^2) = \frac{F_{2D}(x,Q^2) - F_{2D}^{(0)}(x,Q^2)}{F_{2D}^{(0)}(x,Q^2)} \ ,
\label{e:off_D}
\end{equation}
where $F_{2D}^{(0)}$ is calculated by setting $\delta f=0$ or $\delta F$=0. This ratio effectively takes into account all the differences in the computational implementation of nuclear and power corrections by different groups, allowing a cleaner comparison of the detected off-shell effects.

\begin{figure}[b]
    \centering
    \includegraphics[width=0.95\linewidth]{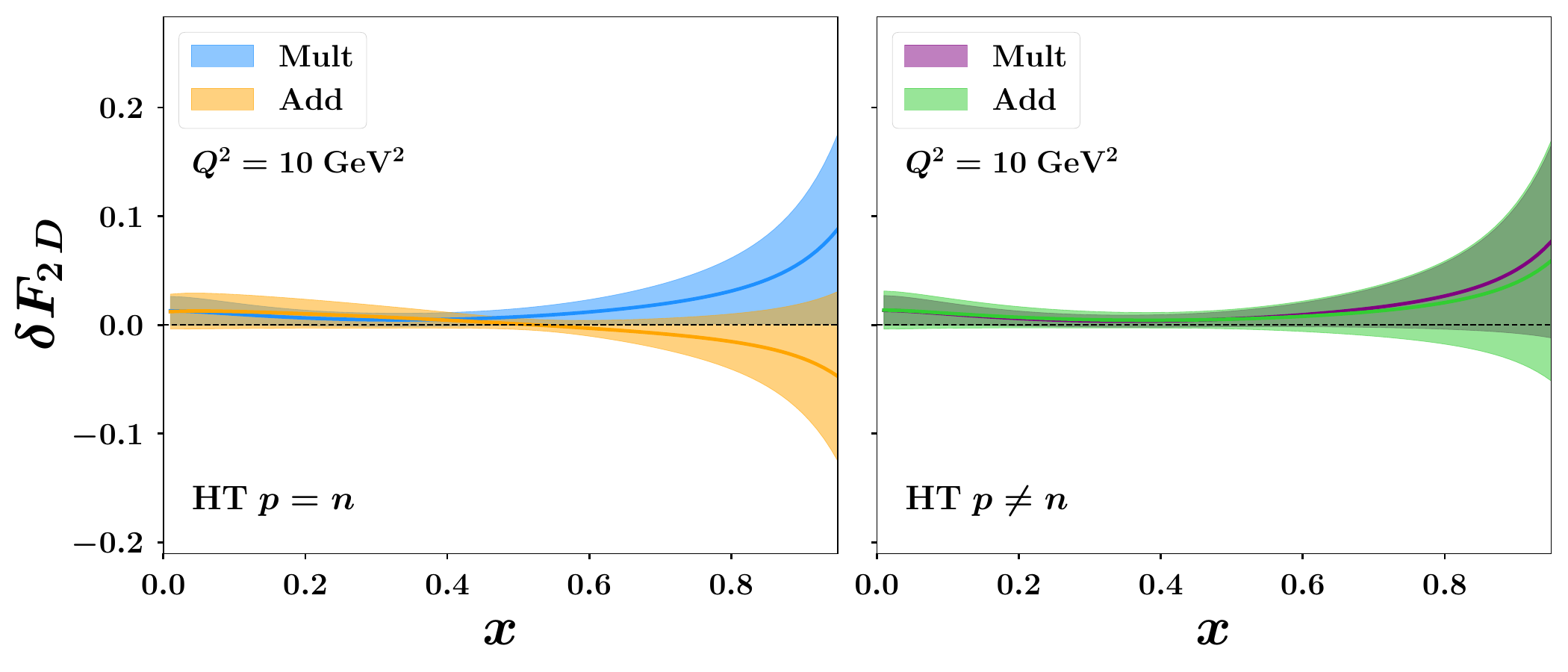}
    \\[-0.6cm]
    \caption{Comparison of the calculated ``effective'' off-shell function $\delta F_{2D}$ obtained with the various HT correction implementations discussed in this paper. \textit{Left panel}: isospin-independent multiplicative (blue band) and additive (orange band) HT corrections (HT $p=n$). \textit{Right panel}: isospin-dependent multiplicative (purple band) and additive (green band) HT corrections (HT $p\neq n$). The bands represent $T^2 = 2.7$ uncertainties.
    }
    \label{f:off_F2_level}
\end{figure}

In Fig.~\ref{f:off_F2_level} we plot the $\delta F_{2D}$ effective off-shell function calculated with the results of our higher-twist fits with isospin-dependent additive (Add) and multiplicative (Mult) implementations. Compared to the off-shell $\delta f$ functions shown in Figs.~\ref{f:Fit_iso} and \ref{f:Fit_noiso}, the effective off-shell function $\delta F_{2D}$ has a smaller excursion away from 0 and larger error bands. The former effect is a direct consequence of deuteron smearing, and the latter is due to the addition of the $d/u$ and HT correction uncertainties in the calculation. It would be interesting to compare these curves with calculations in other fitting frameworks.

\section{$W$-boson vs. decay lepton asymmetries in global QCD fits}
\label{a:doubleC}

In the literature, there has been debate about the reliability of including both the $W \to l + \nu$ charge asymmetry data \cite{D0:2013xqc,D0:2014kma,D0:2013lql} and the reconstructed $W$-boson asymmetry experimental data~\cite{CDF:2005cgc,CDF:2009cjw} from Tevatron in a global QCD analysis. 

In fact, these asymmetries originate from different analyses of the same Tevatron experimental 
events, potentially leading to double-counting of their statistical relevance.
However, in the Tevatron $W$-boson asymmetry measurement, the kinematics of the $W$ boson is reconstructed utilizing the measured missing mass, which is not necessary in the decay lepton measurement. As a result, the reconstructed $W$ asymmetry has a larger information content, which allows it to directly reflect parton-level kinematic, and is sensitive to larger $x$ values than the lepton decay asymmetry.
The lepton asymmetry, on the other hand, depends on the decay kinematics and introduces additional smearing, typically shifting sensitivity toward lower-$x$ values. 

In order to quantify the extent to which these 2 measurements overlap, 
we performed a fit with the same phenomenological setup of the CJ22 analysis~\cite{Accardi:2023gyr}, but excluded the $W$-asymmetry data from the global data set.

\begin{figure}[bth]
\centering
\includegraphics[width=0.97\textwidth]{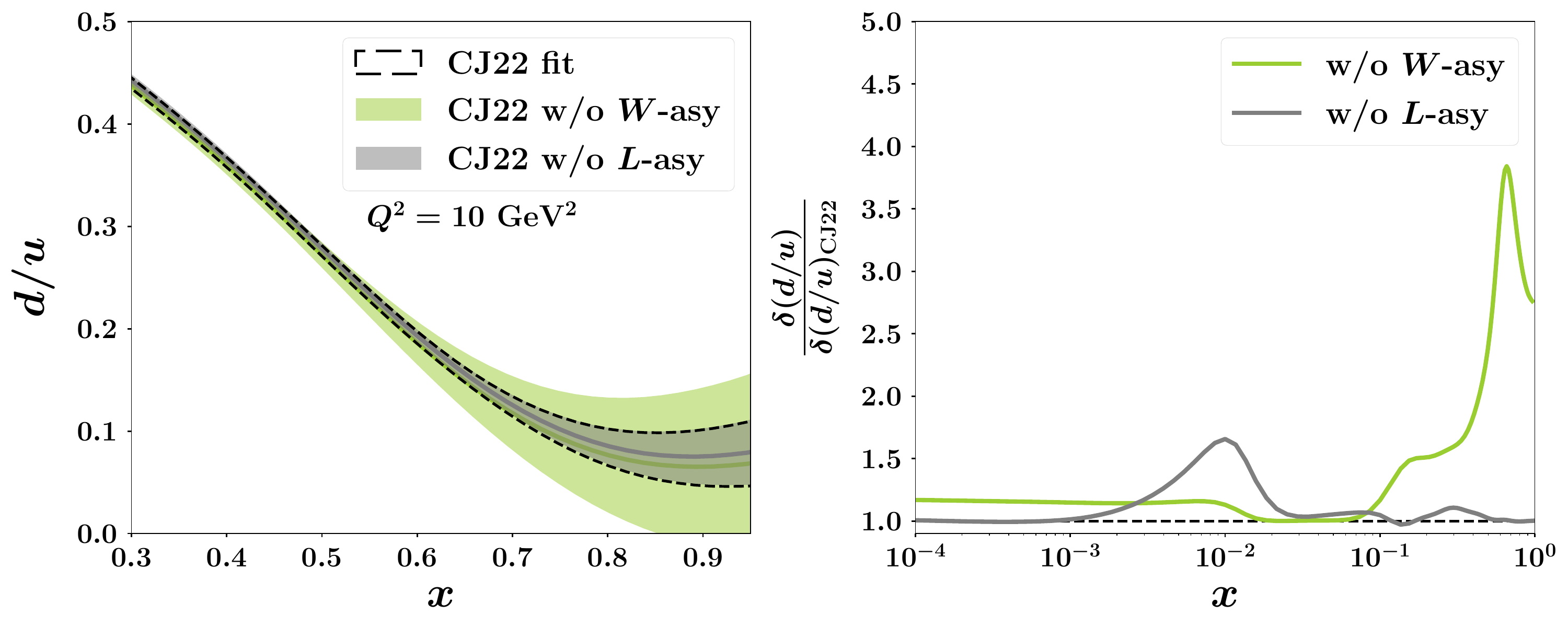}
\vskip-0.3cm
\caption{
Comparison between a standard CJ22 fit (black dashed band) and fit excluding $W$-asymmetry (green band) or lepton-asymmetry (gray band) experimental data. Left panel: $d/u$ PDF ratio. Right panel: ratio between standard CJ22 $d/u$ uncertainties and CJ22 excluding $W$- or lepton-asymmetry data. 
Uncertainty bands represent $T^2 = 2.7$ uncertainties.
}
\label{f:Wasy_impact}
\end{figure}

In Fig.~\ref{f:Wasy_impact}, we compare the extracted $d/u$ PDF ratio from the baseline CJ22 fit (black dashed band) to the fits where either the $W$-asymmetry data (green band) or the lepton-asymmetry data (gray band) are excluded. In the left panel, we observe that the central value of the $d/u$ ratio (dashed black line) remains mostly stable by the inclusion of both data sets. Instead, the error bands are reduced when $W$-asymmetry data are included, particularly in the region where $x > 0.3$ (compare dashed black and green bands). When excluding the lepton-asymmetry data, the difference from the standard CJ22 fit is very small. In the right panel of Fig.~\ref{f:Wasy_impact}, a more detailed analysis of the constraining power of the two data sets is reported. Specifically, we display the $d/u$ uncertainty ratio of the fit excluding the two data sets (separately) to the standard CJ22 one. We note that the bulk of the impact of the lepton-asymmetry data is around $x=0.01$, while that of the $W$-boson asymmetry data is at $x>0.1$. Some small overlap can be seen when $0.001<x<0.01$ and at $x\simeq 0.3$.

In conclusion, the constraining power of the lepton and $W$-boson asymmetries is confined to almost complementary kinematic regions at the partonic level, and the double-counting effect is minimal. Therefore we believe that both data sets can be safely included in global QCD analyses.

\end{document}